\documentclass{aa}

\usepackage[varg]{txfonts}
\usepackage{graphicx}
\usepackage{natbib}
\usepackage{xcolor}
\bibpunct{(}{)}{;}{a}{}{,}

\newcommand\vk{\vec{k}}
\newcommand\vl{\vec{\ell}}
\newcommand\vq{\vec{q}}
\newcommand\vr{\vec{r}}
\newcommand\vs{\vec{s}}
\newcommand\vt{\vec{\theta}}
\newcommand\vx{\vec{x}}
\newcommand\vy{\vec{y}}
\newcommand\hgt{\hat{\gamma_{\mathrm{t}}}}
\newcommand\hds{\hat{\Delta \Sigma}}
\newcommand\hwp{\hat{w_{\mathrm{p}}}}
\newcommand\ho{\hat{\Omega}}
\newcommand\hxi{\hat{\xi}}

\begin{document}

\title{Testing gravity using galaxy-galaxy lensing and clustering
  amplitudes in KiDS-1000, BOSS and 2dFLenS}

\titlerunning{Galaxy-galaxy lensing cosmology in KiDS-1000}

\author{Chris Blake\inst{1}\thanks{E-mail: cblake@swin.edu.au}
\and Alexandra Amon\inst{2}
\and Marika Asgari\inst{3}
\and Maciej Bilicki\inst{4}
\and Andrej Dvornik\inst{5}
\and Thomas Erben\inst{6}
\and Benjamin Giblin\inst{3}
\and Karl Glazebrook\inst{1}
\and Catherine Heymans\inst{3,5}
\and Hendrik Hildebrandt\inst{5}
\and Benjamin Joachimi\inst{7}
\and Shahab Joudaki\inst{8}
\and Arun Kannawadi\inst{9,10}
\and Konrad Kuijken\inst{10}
\and Chris Lidman\inst{11}
\and David Parkinson\inst{12}
\and HuanYuan Shan\inst{13,14}
\and Tilman Tr\"oster\inst{3}
\and Jan Luca van den Busch\inst{5}
\and Christian Wolf\inst{11}
\and Angus H.\ Wright\inst{5}
}

\authorrunning{Blake et al.}

\institute{Centre for Astrophysics \& Supercomputing, Swinburne
  University of Technology, P.O.\ Box 218, Hawthorn, VIC 3122,
  Australia
\and Kavli Institute for Particle Astrophysics \& Cosmology, P.O.\ Box
2450, Stanford University, Stanford, CA 94305, USA
\and Institute for Astronomy, University of Edinburgh, Royal
Observatory, Blackford Hill, Edinburgh, EH9 3HJ, UK
\and Center for Theoretical Physics, Polish Academy of Sciences,
al. Lotnik\'ow 32/46, 02-668, Warsaw, Poland
\and Ruhr-University Bochum, Astronomical Institute, German Centre for
Cosmological Lensing, Universit\"atsstr. 150, 44801 Bochum, Germany
\and Argelander-Institut f\"ur Astronomie, Auf dem H\"ugel 71, 53121
Bonn, Germany
\and Department of Physics and Astronomy, University College London,
Gower Street, London WC1E 6BT, UK
\and Department of Physics, University of Oxford, Denys Wilkinson
Building, Keble Road, Oxford OX1 3RH, UK
\and Department of Astrophysical Sciences, Princeton University, 4 Ivy
Lane, Princeton, NJ 08544, USA
\and Leiden Observatory, Leiden University, P.O.Box 9513, 2300RA
Leiden, The Netherlands
\and Research School of Astronomy and Astrophysics, Australian
National University, Canberra ACT 2611, Australia
\and Korea Astronomy and Space Science Institute, 776 Daedeokdae-ro,
Yuseong-gu, Daejeon 34055, Republic of Korea
\and Shanghai Astronomical Observatory (SHAO), Nandan Road 80,
Shanghai 200030, China
\and University of Chinese Academy of Sciences, Beijing 100049, China
}

\date{Received date / Accepted date}

\abstract{The physics of gravity on cosmological scales affects both
  the rate of assembly of large-scale structure and the gravitational
  lensing of background light through this cosmic web.  By comparing
  the amplitude of these different observational signatures, we can
  construct tests that can distinguish general relativity from its
  potential modifications.  We used the latest weak gravitational
  lensing dataset from the Kilo-Degree Survey, KiDS-1000, in
  conjunction with overlapping galaxy spectroscopic redshift surveys,
  BOSS and 2dFLenS, to perform the most precise existing
  amplitude-ratio test.  We measured the associated $E_{\mathrm{G}}$
  statistic with $15-20\%$ errors in five $\Delta z = 0.1$
  tomographic redshift bins in the range $0.2 < z < 0.7$ on projected
  scales up to $100 \, h^{-1}$ Mpc.  The scale-independence and
  redshift-dependence of these measurements are consistent with the
  theoretical expectation of general relativity in a Universe with
  matter density $\Omega_{\mathrm{m}} = 0.27 \pm 0.04$.  We
  demonstrate that our results are robust against different analysis
  choices, including schemes for correcting the effects of source
  photometric redshift errors, and we compare the performance of
  angular and projected galaxy-galaxy lensing statistics.}


\maketitle

\section{Introduction}
\label{secintro}

A central goal of modern cosmology is to discover whether the dark
energy that appears to fill the Universe is associated with its
matter-energy content, laws of gravity, or some alternative physics.
A compelling means of distinguishing between these scenarios is to
analyse the different observational signatures that are present in the
clumpy, inhomogeneous Universe, which powerfully complements
measurements of the expansion history of the smooth, homogeneous
Universe
\citep[e.g.][]{Linder05,Wang08,Guzzo08,Weinberg13,Huterer15}.

Two important observational probes of the inhomogeneous Universe are
the peculiar velocities induced in galaxies by the gravitational
collapse of large-scale structure, which are statistically imprinted
in galaxy redshift surveys as redshift-space distortions
\citep[e.g.][]{Hamilton98,Scoccimarro04,Song09}, and the gravitational
lensing of light by the cosmic web, which may be measured using cosmic
shear surveys \citep[e.g.][]{Bartelmann01,Kilbinger15,Mandelbaum18}.
These probes are complementary because they allow for the
differentiation between the two space-time metric potentials which
govern the motion of non-relativistic particles, such as galaxy
tracers, and the gravitational deflection of light.  The difference or
`gravitational slip' between these potentials is predicted to be zero
in general relativity, but it may be significant in modified gravity
scenarios
\citep[e.g.][]{Uzan01,Zhang07,Jain10,Bertschinger11,Clifton12}.

Recent advances in weak gravitational lensing datasets, including the
Kilo-Degree Survey \citep[KiDS,][]{Hildebrandt20}, the Dark Energy
Survey \citep[DES,][]{Abbott18} and the Subaru Hyper Suprime-Cam
Survey \cite[HSC,][]{Hikage19}, have led to dramatic improvements in
the quality of these observational tests.  Gravitational lensing now
permits the accurate determination of, and combinations of, important
cosmological parameters such as the matter density of the Universe and
normalisation of the matter power spectrum, and thereby detailed
comparisons with other cosmological probes such as galaxy clustering
\citep{Alam17b} and the cosmic microwave background radiation
\citep{Planck18}.  Some of these comparisons have yielded intriguing
evidence of `tension' on both small and large scales
\citep[e.g.][]{Joudaki17,Leauthaud17,Lange19,Hildebrandt20,Asgari20a},
which are currently unresolved.

In this paper we perform a new study regarding this question using the
latest weak gravitational lensing dataset from the Kilo-Degree Survey,
KiDS-1000 \citep{Kuijken19}, in conjunction with overlapping galaxy
spectroscopic redshift survey data from the Baryon Oscillation
Spectroscopic Survey \citep[BOSS,][]{Reid16} and the 2-degree Field
Lensing Survey \citep[2dFLenS,][]{Blake16b}.  In particular, we focus
on a simple implementation of the lensing-clustering test which
compares the amplitude of gravitational lensing around foreground
galaxies (commonly known as galaxy-galaxy lensing), tracing
low-redshift overdensities, with the amplitude of galaxy velocities
induced by these overdensities and measured by redshift-space
distortions, which constitutes an amplitude-ratio test.  This
diagnostic was first proposed by \citet{Zhang07} as the
$E_{\mathrm{G}}$ statistic and implemented in its current form by
\citet{Reyes10} using data from the Sloan Digital Sky Survey.  These
measurements have subsequently been refined by a series of studies
\citep{Blake16a,Pullen16,Alam17a,delaTorre17,Amon18,Singh19,Jullo19}
which have used new datasets to increase the accuracy of the
amplitude-ratio determination, albeit showing some evidence of
internal disagreement.

The availability of the KiDS-1000 dataset and associated calibration
samples has allowed us to perform the most accurate existing
amplitude-ratio test, on projected scales up to $100 \, h^{-1}$ Mpc,
including rigorous systematic-error control.  As part of this
analysis, we use these datasets and representative simulations to
study the efficacy of different corrections for the effects of source
photometric redshift errors, comparing different galaxy-galaxy lensing
estimators and the relative performance of angular and projected
statistics.  Our analysis sets the stage for future per-cent level
implementations of these tests using new datasets from the Dark Energy
Spectroscopic Instrument \citep[DESI,][]{DESI16}, the 4-metre
Multi-Object Spectrograph Telescope \citep[4MOST,][]{deJong19}, the
Rubin Observatory Legacy Survey of Space and Time
\citep[LSST,][]{Ivezic19}, and the {\it Euclid} satellite
\citep{Laureijs11}.

This paper is structured as follows: in Sect.\ \ref{sectheory} we
review the theoretical correlations between weak lensing and
overdensity observables, on which galaxy-galaxy lensing studies are
based.  In Sect.\ \ref{secest} we summarise the angular and projected
galaxy-galaxy lensing estimators derived from these correlations, with
particular attention to the effect of source photometric redshift
errors.  In Sect.\ \ref{secamp} we introduce the amplitude-ratio test
between galaxy-galaxy lensing and clustering observables, constructed
from annular differential surface density statistics, and in
Sect.\ \ref{seccov} we derive the analytical covariances of these
estimators in the Gaussian approximation, including the effects of the
survey window function.  We introduce the KiDS-1000 weak lensing and
overlapping Luminous Red Galaxy (LRG) spectroscopic datasets in
Sect.\ \ref{secdata}.  We create representative survey mock catalogues
in Sect.\ \ref{secmocks}, which we use to verify our cosmological
analysis in Sect.\ \ref{secmocktests}.  Finally, we describe the
results of our cosmological tests applied to the KiDS-LRG datasets in
Sect.\ \ref{secdatatests}.  We summarise our investigation in
Sect.\ \ref{secsummary}.

\section{Theory}
\label{sectheory}

In this section we briefly review the theoretical expressions for the
auto- and cross-correlations between weak gravitational lensing and
galaxy overdensity observables, which form the basis of galaxy-galaxy
lensing studies.

\subsection{Lensing convergence and tangential shear}

The observable effects of weak gravitational lensing, on a source
located at co-moving co-ordinate $\chi_{\mathrm{s}}$ in sky direction
$\ho$, can be expressed in terms of the lensing convergence $\kappa$
\citep[for reviews, see][]{Bartelmann01,Kilbinger15,Mandelbaum18}.
The convergence is a weighted integral over co-moving distance $\chi$
of the matter overdensity $\delta_{\mathrm{m}}$ along the
line-of-sight, which we can write as
\begin{equation}
  \kappa(\chi_{\mathrm{s}},\ho) = \frac{3 \Omega_{\mathrm{m}} H_0^2}{2
    c^2} \int_0^{\chi_{\mathrm{s}}} d\chi \, \frac{\chi \,
    (\chi_{\mathrm{s}} - \chi)}{\chi_{\mathrm{s}}} \,
  \frac{\delta_{\mathrm{m}}(\chi,\ho)}{a(\chi)} ,
\label{eqkappa1}
\end{equation}
assuming (throughout this paper) a spatially-flat Universe, where
$\Omega_{\mathrm{m}}$ is the matter density as a fraction of the
critical density, $H_0$ is the Hubble parameter, $c$ is the speed of
light, and $a = 1/(1+z)$ is the cosmic scale factor at redshift $z$.
We can conveniently write Eq.\ \ref{eqkappa1} in terms of the critical
surface mass density at a lens plane at co-moving distance
$\chi_{\mathrm{l}}$,
\begin{equation}
  \Sigma_{\mathrm{c}}(\chi_{\mathrm{l}},\chi_{\mathrm{s}}) =
  \frac{c^2}{4 \mathrm{\pi} G}
  \frac{\chi_{\mathrm{s}}}{(1+z_{\mathrm{l}}) \, \chi_{\mathrm{l}} \,
    (\chi_{\mathrm{s}} - \chi_{\mathrm{l}})} ,
\end{equation}
where $G$ is the gravitational constant, and $\chi_{\mathrm{s}} >
\chi_{\mathrm{l}}$.  Hence,
\begin{equation}
  \kappa(\chi_{\mathrm{s}},\ho) = \overline{\rho_{\mathrm{m}}}
  \int_0^{\chi_{\mathrm{s}}} d\chi \,
  \Sigma_{\mathrm{c}}^{-1}(\chi,\chi_{\mathrm{s}}) \,
  \delta_{\mathrm{m}}(\chi,\ho) ,
\label{eqkappa2}
\end{equation}
where $\overline{\rho_{\mathrm{m}}}$ is the mean matter
density.\footnote{We refer the reader to \citet{Dvornik18} Appendix C
  for a full discussion of the different definitions of
  $\Sigma_{\mathrm{c}}$ that have been adopted in the literature.}

Suppose that the overdensity is associated with an isolated lens
galaxy at distance $\chi_{\mathrm{l}}$ in an otherwise homogeneous
Universe.  In this case, Eq.\ \ref{eqkappa2} may be written in the
form
\begin{equation}
  \kappa(\chi_{\mathrm{l}},\chi_{\mathrm{s}},\ho) \approx
  \overline{\rho_{\mathrm{m}}} \,
  \Sigma_{\mathrm{c}}^{-1}(\chi_{\mathrm{l}},\chi_{\mathrm{s}}) \,
  \int_0^{\chi_{\mathrm{s}}} d\chi \, \delta_{\mathrm{m}}(\chi,\ho) .
\label{eqkappa3}
\end{equation}
Eq.\ \ref{eqkappa3} motivates that the weak lensing observable can be
related to the projected mass density around the lens, $\Sigma = \int
\rho_{\mathrm{m}} \, d\chi$, where $\delta_{\mathrm{m}} =
\rho_{\mathrm{m}}/\overline{\rho_{\mathrm{m}}} - 1$.  The convergence
may be written in terms of this quantity as
\begin{equation}
  \kappa(\chi_{\mathrm{l}},\chi_{\mathrm{s}},\ho) \approx
  \Sigma_{\mathrm{c}}^{-1}(\chi_{\mathrm{l}},\chi_{\mathrm{s}}) \left(
  \Sigma - \overline{\Sigma} \right) ,
\end{equation}
where $\overline{\Sigma} = \int \overline{\rho} \, d\chi$ represents
the average background, emphasising that gravitational lensing traces
the increment between the mass density and the background.

The average tangential shear $\gamma_{\mathrm{t}}$ at angular
separation $\theta$ from an axisymmetric lens is related to the
convergence as
\begin{equation}
  \langle \gamma_{\mathrm{t}}(\theta) \rangle = \langle
  \overline{\kappa}(<\theta) \rangle - \langle \kappa(\theta) \rangle
  ,
\label{eqgt}
\end{equation}
where $\overline{\kappa}(<\theta)$ is the mean convergence within
separation $\theta$.  At the location of the lens, angular separations
are related to projected separations as $R = \chi(z_{\mathrm{l}}) \,
\theta$.  Defining the differential projected surface mass density
around the lens as a function of projected separation,
\begin{equation}
\Delta \Sigma(R) = \overline{\Sigma}(<R) - \Sigma(R) ,
\label{eqdsigdef}
\end{equation}
where,
\begin{equation}
\overline{\Sigma}(<R) = \frac{2}{R^2} \int_0^R R' \, \Sigma(R') \, dR'
,
\end{equation}
we find that for a single source-lens pair at distances
$\chi_{\mathrm{l}}$ and $\chi_{\mathrm{s}}$ (omitting the angled
brackets),
\begin{equation}
  \gamma_{\mathrm{t}}(\theta) =
  \Sigma_{\mathrm{c}}^{-1}(\chi_{\mathrm{l}},\chi_{\mathrm{s}}) \,
  \Delta\Sigma(R) .
\label{eqgtdsig}
\end{equation}

\subsection{Galaxy-convergence cross-correlation}

For an ensemble of sources with distance probability distribution
$p_{\mathrm{s}}(\chi)$ (normalised such that $\int
p_{\mathrm{s}}(\chi) \, d\chi = 1$), the total convergence in a given
sky direction is
\begin{equation}
\begin{split}
  \kappa(\ho) &= \int d\chi_{\mathrm{s}} \,
  p_{\mathrm{s}}(\chi_{\mathrm{s}}) \, \kappa(\chi_{\mathrm{s}},\ho)
  \\ &= \overline{\rho_{\mathrm{m}}} \int_0^\infty d\chi \,
  \overline{\Sigma_{\mathrm{c}}^{-1}}(\chi) \,
  \delta_{\mathrm{m}}(\chi,\ho) ,
\end{split}
\end{equation}
where,
\begin{equation}
\overline{\Sigma_{\mathrm{c}}^{-1}}(\chi) = \int_\chi^\infty
d\chi_{\mathrm{s}} \, p_{\mathrm{s}}(\chi_{\mathrm{s}}) \,
\Sigma_{\mathrm{c}}^{-1}(\chi,\chi_{\mathrm{s}}) ,
\label{eqavesigc}
\end{equation}
with the lower limit of the integral applying because
$\Sigma_{\mathrm{c}}^{-1}(\chi_{\mathrm{l}},\chi_{\mathrm{s}}) = 0$
for $\chi_{\mathrm{s}} < \chi_{\mathrm{l}}$.  We consider forming the
angular cross-correlation function of this convergence field with the
projected number overdensity of an ensemble of lenses with distance
probability distribution $p_{\mathrm{l}}(\chi)$,
\begin{equation}
  \delta_{\mathrm{g,2D}}(\ho) = \int d\chi \, p_{\mathrm{l}}(\chi) \,
  \delta_{\mathrm{g}}(\chi,\ho) .
\end{equation}
The galaxy-convergence cross-correlation function at angular
separation $\vt$ is,
\begin{equation}
\omega_{\mathrm{g\kappa}}(\vt) = \langle \kappa(\ho) \,
\delta_{\mathrm{g,2D}}(\ho + \vt) \rangle .
\end{equation}
Expressing the overdensity fields in terms of their Fourier components
we find, after some algebra,
\begin{equation}
 \omega_{\mathrm{g\kappa}}(\vt) = \int \frac{d^2\vl}{(2\mathrm{\pi})^2} \,
 C_{\mathrm{g\kappa}}(\vl) \, \mathrm{e}^{-\mathrm{i}\vl\cdot\vt} ,
\label{eqwgk}
\end{equation}
where $\vl$ is a 2D Fourier wavevector, and the corresponding angular
cross-power spectrum $C_{\mathrm{g\kappa}}(\ell)$ is given by
\citep{Guzik01,Hu04,Joachimi10},
\begin{equation}
C_{\mathrm{g\kappa}}(\ell) = \overline{\rho_{\mathrm{m}}} \int d\chi
\, p_{\mathrm{l}}(\chi) \,
\frac{\overline{\Sigma_{\mathrm{c}}^{-1}}(\chi)}{\chi^2} \,
P_{\mathrm{gm}} \left( \frac{\ell}{\chi},\chi \right) ,
\label{eqclgk}
\end{equation}
where $P_{\mathrm{gm}}(k,\chi)$ is the 3D galaxy-matter cross-power
spectrum at wavenumber $k$ and distance $\chi$.  Taking the azimuthal
average of Eq.\ \ref{eqwgk} over all directions $\vt$, the complex
exponential integrates to a Bessel function of the first kind,
$J_0(x)$, such that,
\begin{equation}
\omega_{\mathrm{g\kappa}}(\theta) = \int
\frac{d^2\vl}{(2\mathrm{\pi})^2} \, C_{\mathrm{g\kappa}}(\vl) \,
J_0(\ell \theta) = \int \frac{d\ell \, \ell}{2\mathrm{\pi}} \,
C_{\mathrm{g\kappa}}(\ell) \, J_0(\ell \theta) .
\end{equation}
Using Eq.\ \ref{eqgt} and Bessel function identities, we can then
obtain an expression for the statistical average tangential shear
around an ensemble of lenses,
\begin{equation}
\gamma_{\mathrm{t}}(\theta) =
\overline{\omega_{\mathrm{g\kappa}}}(<\theta) -
\omega_{\mathrm{g\kappa}}(\theta) = \int \frac{d\ell \,
  \ell}{2\mathrm{\pi}} \, C_{\mathrm{g\kappa}}(\ell) \, J_2(\ell
\theta) .
\label{eqgtmod}
\end{equation}
Likewise, we can generalise Eq.\ \ref{eqgtdsig} to apply to broad
source and lens distributions:
\begin{equation}
\gamma_{\mathrm{t}}(\theta) = \int d\chi \, p_{\mathrm{l}}(\chi) \,
\overline{\Sigma_{\mathrm{c}}^{-1}}(\chi) \, \Delta\Sigma(R,\chi) .
\label{eqgtdsigbroad}
\end{equation}
Comparing the formulations of Eqs.\ \ref{eqgtmod} and
\ref{eqgtdsigbroad} allows us to demonstrate that,
\begin{equation}
  \Sigma(R) = \overline{\rho_{\mathrm{m}}} \int_{-\infty}^\infty d\Pi
  \, \left[ 1 + \xi_{\mathrm{gm}}(R,\Pi) \right] ,
\label{eqsig}
\end{equation}
in terms of the 3D galaxy-matter cross-correlation function
$\xi_{\mathrm{gm}}(R,\Pi)$ at projected separation $R$ and
line-of-sight separation $\Pi$, where the constant term `$1+$'
cancels out in the evaluation of the observable $\Delta \Sigma$.
After some algebra we find,
\begin{equation}
\Delta \Sigma(R) = \overline{\rho_{\mathrm{m}}} \int_0^\infty dr \,
W(r,R) \, \xi_{\mathrm{gm}}(r) ,
\end{equation}
where,
\begin{equation}
  W(r,R) = \frac{4r^2}{R^2} - \left[ \frac{4r \sqrt{r^2-R^2}}{R^2} +
    \frac{2r}{\sqrt{r^2 - R^2}} \right] \, H(r-R) ,
\end{equation}
where $H(x) = 0$ if $x<0$ and $H(x) = 1$ if $x>0$ is the Heaviside
step function.  The relations in this section make the approximations
of using the Limber equation \citep{Limber53} and neglecting
additional effects such as cosmic magnification \citep{Unruh20} and
intrinsic alignments \citep{Joachimi15}.

\subsection{Auto-correlation functions}

In order to determine the analytical covariance in
Sect.\ \ref{seccov}, we also need expressions for the auto-correlation
functions of the convergence, $\omega_{\mathrm{\kappa\kappa}}(\vt) =
\int \frac{d^2\vl}{(2\mathrm{\pi})^2} \,
C_{\mathrm{\kappa\kappa}}(\vl) \,
\mathrm{e}^{-\mathrm{i}\vl\cdot\vt}$, and the galaxy overdensity,
$\omega_{\mathrm{gg}}(\vt) = \int \frac{d^2\vl}{(2\mathrm{\pi})^2} \,
C_{\mathrm{gg}}(\vl) \, \mathrm{e}^{-\mathrm{i}\vl\cdot\vt}$.  Given
two source populations with distance probability distributions
$p_{\mathrm{s},1}(\chi)$ and $p_{\mathrm{s},2}(\chi)$, and associated
integrated critical density functions
$\overline{\Sigma^{-1}_{\mathrm{c},1}}$ and
$\overline{\Sigma^{-1}_{\mathrm{c},2}}$, the angular power spectrum of
the convergence is given by,
\begin{equation}
  C_{\mathrm{\kappa\kappa}}(\ell) = \overline{\rho_{\mathrm{m}}}^2
  \int d\chi \, \frac{\overline{\Sigma_{\mathrm{c},1}^{-1}}(\chi) \,
    \overline{\Sigma_{\mathrm{c},2}^{-1}}(\chi)}{\chi^2} \,
  P_{\mathrm{mm}}\left( \frac{\ell}{\chi},\chi \right) ,
\end{equation}
where $P_{\mathrm{mm}}(k,\chi)$ is the 3D (non-linear) matter power
spectrum at wavenumber $k$ and distance $\chi$.  Likewise, for two
projected galaxy overdensity fields with distance probability
distributions $p_{\mathrm{l},1}(\chi)$ and $p_{\mathrm{l},2}(\chi)$,
the angular power spectrum is,
\begin{equation}
C_{\mathrm{gg}}(\ell) = \int d\chi \, \frac{p_{\mathrm{l},1}(\chi) \,
  p_{\mathrm{l},2}(\chi)}{\chi^2} \, P_{\mathrm{gg}}\left(
\frac{\ell}{\chi},\chi \right) ,
\label{eqclgg}
\end{equation}
where $P_{\mathrm{gg}}(k,\chi)$ is the 3D galaxy power spectrum.

\subsection{Bias model}
\label{secbias}

We computed the linear matter power spectrum $P_{\mathrm{L}}(k)$ in
our models using the CAMB software package \citep{Lewis00}, and
evaluated the non-linear matter power spectrum $P_{\mathrm{mm}}(k)$
including the `halofit' corrections \citep[][we define the fiducial
  cosmological parameters used for the simulation and data analysis in
  subsequent sections]{Smith03,Takahashi12}.  We adopted a model for
the non-linear galaxy-galaxy and galaxy-matter 2-point functions,
appearing in Eqs.\ \ref{eqclgk} and \ref{eqclgg}, following
\citet{Baldauf10} and \citet{Mandelbaum13}.  This model assumes a
local, non-linear galaxy bias relation via a Taylor expansion of the
galaxy density field in terms of the matter overdensity,
$\delta_{\mathrm{g}} = b_{\mathrm{L}} \, \delta_{\mathrm{m}} +
\frac{1}{2} b_{\mathrm{NL}} \, \delta_{\mathrm{m}}^2 + ...$, defining
a linear bias parameter $b_{\mathrm{L}}$ and non-linear bias parameter
$b_{\mathrm{NL}}$.  The auto- and cross-correlation statistics in this
model can be written in the form \citep{McDonald06,Smith09},
\begin{equation}
  \begin{split}
    \xi_{\mathrm{gg}} &= b_{\mathrm{L}}^2 \, \xi_{\mathrm{mm}} + 2 \,
    b_{\mathrm{L}} \, b_{\mathrm{NL}} \, \xi_{\mathrm{A}} +
    \frac{1}{2} \, b_{\mathrm{NL}}^2 \, \xi_{\mathrm{B}} ,
    \\ \xi_{\mathrm{gm}} &= b_{\mathrm{L}} \, \xi_{\mathrm{mm}} +
    b_{\mathrm{NL}} \, \xi_{\mathrm{A}} ,
  \end{split}
\end{equation}
where $\xi_{\mathrm{mm}}$ is the correlation function corresponding to
$P_{\mathrm{mm}}(k)$, and $\xi_{\mathrm{A}}$ and $\xi_{\mathrm{B}}$
are obtained by computing the Fourier transforms of,
\begin{equation}
  \begin{split}
    A(k) &= \int \frac{d^3\vq}{(2\mathrm{\pi})^3} \, F_2(\vq,\vk-\vq)
    \, P_{\mathrm{L}}(q) \, P_{\mathrm{L}}(|\vk-\vq|) , \\ B(k) &=
    \int \frac{d^3\vq}{(2\mathrm{\pi})^3} \, P_{\mathrm{L}}(q) \,
    P_{\mathrm{L}}(|\vk-\vq|) ,
  \end{split}
\end{equation}
which depend on the mode-coupling kernel in standard perturbation
theory,
\begin{equation}
  F_2(\vq_1,\vq_2) = \frac{5}{7} + \frac{1}{2} \frac{\vq_1
    . \vq_2}{q_1 q_2} \left( \frac{q_1}{q_2} + \frac{q_2}{q_1} \right)
  + \frac{2}{7} \left( \frac{\vq_1 . \vq_2}{q_1 q_2} \right)^2 .
\end{equation}
We evaluated these integrals using the {\tt FAST} software package
\citep{McEwen16} and note that $\xi_{\mathrm{B}} =
\xi_{\mathrm{L}}^2$, where $\xi_{\mathrm{L}}$ is the correlation
function corresponding to $P_{\mathrm{L}}(k)$.  This model is only
expected to be valid on scales exceeding the virial radius of dark
matter haloes, since it does not address halo exclusion, the
distribution of galaxies within haloes, or other forms of stochastic
or non-local effects \citep{Asgari20b}.  However, this 2-parameter
bias model is adequate for our large-scale analysis, which we verify
using representative mock catalogues in Sect.\ \ref{secmocktests}.

\section{Estimators}
\label{secest}

In this section we specify estimators that may be used to measure
$\gamma_{\mathrm{t}}(\theta)$ and $\Delta \Sigma(R)$ from ensembles of
sources and lenses, and discuss how estimates of $\Delta \Sigma(R)$
are affected by uncertainties in source distances.

\subsection{Average tangential shear $\gamma_{\mathrm{t}}(\theta)$}

We can estimate the average tangential shear of a set of sources (s)
around lenses (l) by evaluating the following expression
\citep{Mandelbaum06}, which also utilises an unclustered random lens
catalogue (r) with the same selection function as the lenses:
\begin{equation}
\hgt(\theta) = \frac{\sum\limits_{\mathrm{ls}} w_{\mathrm{l}} \,
  w_{\mathrm{s}} \, e_{\mathrm{t,ls}} - \sum\limits_{\mathrm{rs}}
  w_{\mathrm{r}} \, w_{\mathrm{s}} \,
  e_{\mathrm{t,rs}}}{\sum\limits_{\mathrm{rs}} w_{\mathrm{r}} \,
  w_{\mathrm{s}}} .
\label{eqgtest}
\end{equation}
The sums in Eq.\ \ref{eqgtest} are taken over pairs of sources and
lenses with angular separations within a bin around $\theta$, $w_i$
are weights applied to the different samples (normalised such that
$\sum_{\mathrm{l}} w_{\mathrm{l}} = \sum_{\mathrm{r}}
w_{\mathrm{r}}$), and $e_{\mathrm{t}}$ indicates the tangential
ellipticity of the source, projected onto an axis normal to the line
joining the source and lens (or random lens).

Eq.\ \ref{eqgtest} involves the random lens catalogue in two places.
First, the tangential shear of sources around random lenses is
subtracted from the data signal.  The subtracted term has an
expectation value of zero, but significantly decreases the variance of
the estimator at large separations \citep{Singh17}.  Second, the
estimator is normalised by a sum over pairs of sources and random
lenses, rather than data lenses.  This ensures that the estimator is
unbiased: the alternative estimator, $\hgt = \sum_{\mathrm{ls}}
w_{\mathrm{l}} \, w_{\mathrm{s}} \, e_{\mathrm{t,ls}} /
\sum_{\mathrm{ls}} w_{\mathrm{l}} \, w_{\mathrm{s}}$, is biased by any
source-lens clustering (if the angular cross-correlation function
$\omega_{\mathrm{ls}}(\theta) \ne 0$), which would modify the
denominator of the expression but not the numerator.  The magnitude of
this effect is sometimes known as the `boost' factor
\citep{Sheldon04},
\begin{equation}
  B(\theta) = \frac{\sum\limits_{\mathrm{ls}} w_{\mathrm{l}} \,
    w_{\mathrm{s}}}{\sum\limits_{\mathrm{rs}} w_{\mathrm{r}} \,
    w_{\mathrm{s}}} ,
\end{equation}
where the sums are again taken over source-lens pairs with angular
separations within a given bin.  We note that $\langle B(\theta)
\rangle = 1 + \omega_{\mathrm{ls}}(\theta)$ for unity weights.

\subsection{Projected mass density $\Delta\Sigma(R)$}

Assuming the source and lens distances are known, each source-lens
pair may be used to estimate the projected mass density around the
lenses by inverting Eq.\ \ref{eqgtdsig}:
\begin{equation}
  \hds(R) = e_{\mathrm{t}}(R/\chi_{\mathrm{l}}) \,
  \Sigma_{\mathrm{c}}(\chi_{\mathrm{l}},\chi_{\mathrm{s}}) .
\end{equation}
For an ensemble of sources and lenses, the mean projected mass density
may then be estimated by an expression analogous to Eq.\
\ref{eqgtest} \citep{Singh17},
\begin{equation}
\begin{split}
  \hds(R) = \frac{\sum\limits_{\mathrm{ls}} w_{\mathrm{l}} \, w_{\mathrm{s}}
    \, w_{\mathrm{ls}} \, e_{\mathrm{t,ls}}(R/\chi_{\mathrm{l}}) \,
    \Sigma_{\mathrm{c}}(\chi_{\mathrm{l}},\chi_{\mathrm{s}})}{\sum\limits_{\mathrm{rs}}
    w_{\mathrm{r}} \, w_{\mathrm{s}} \, w_{\mathrm{rs}}} \\ -
  \frac{\sum\limits_{\mathrm{rs}} w_{\mathrm{r}} \, w_{\mathrm{s}} \,
    w_{\mathrm{rs}} \, e_{\mathrm{t,rs}}(R/\chi_{\mathrm{r}}) \,
    \Sigma_{\mathrm{c}}(\chi_{\mathrm{r}},\chi_{\mathrm{s}})}{\sum\limits_{\mathrm{rs}}
    w_{\mathrm{r}} \, w_{\mathrm{s}} \, w_{\mathrm{rs}}} ,
\end{split}
\label{eqdsigest1}
\end{equation}
where we have allowed for an additional pair weight between sources
and lenses, $w_{\mathrm{ls}}$, and random lenses, $w_{\mathrm{rs}}$.
Assuming a constant shape noise in $e_{\mathrm{t}}$, the noise in the
estimate of $\Delta \Sigma(R) = e_{\mathrm{t}} \, \Sigma_{\mathrm{c}}$
from each source-lens pair is proportional to $\Sigma_{\mathrm{c}}$,
hence the optimal inverse-variance weight is $w_{\mathrm{ls}} \propto
\Sigma_{\mathrm{c}}^{-2}$, and the weighted estimator may be written,
\begin{equation}
\begin{split}
  \hds(R) = \frac{\sum\limits_{\mathrm{ls}} w_{\mathrm{l}} \, w_{\mathrm{s}}
    \, e_{\mathrm{t,ls}}(R/\chi_{\mathrm{l}}) \,
    \Sigma^{-1}_{\mathrm{c,ls}}}{\sum\limits_{\mathrm{rs}} w_{\mathrm{r}} \,
    w_{\mathrm{s}} \, \Sigma^{-2}_{\mathrm{c,rs}}} \\ -
  \frac{\sum\limits_{\mathrm{rs}} w_{\mathrm{r}} \, w_{\mathrm{s}} \,
    e_{\mathrm{t,rs}}(R/\chi_{\mathrm{r}}) \,
    \Sigma^{-1}_{\mathrm{c,rs}}}{\sum\limits_{\mathrm{rs}} w_{\mathrm{r}} \,
    w_{\mathrm{s}} \, \Sigma^{-2}_{\mathrm{c,rs}}} .
\end{split}
\label{eqdsigest2}
\end{equation}

\subsection{Photo-$z$ dilution correction for $\Delta\Sigma(R)$}
\label{secphotoz}

The difficulty faced when determining $\Delta \Sigma$ is that source
distances are typically only accessible through photometric redshifts
and may contain significant errors, leading to a bias in the estimate
through incorrect scaling factors $\Sigma_{\mathrm{c}}$ (we assume in
this discussion that spectroscopic lens distances are available).  For
example, sources may apparently lie behind lenses according to their
photometric redshift, whilst in fact being positioned in front of the
lenses and contributing no galaxy-galaxy lensing signal, creating a
downward bias in the measurement.

For a single source-lens pair, the estimated value of
$\Sigma_{\mathrm{c}}$ for the pair based on the source photometric
redshift, $\Sigma_{\mathrm{c,lp}}$, may differ from its true value
based on the source spectroscopic redshift, $\Sigma_{\mathrm{c,ls}}$,
\begin{equation}
\hds = e_{\mathrm{t}} \, \Sigma_{\mathrm{c,lp}} = \left( \frac{\Delta
  \Sigma^{\mathrm{true}}}{\Sigma_{\mathrm{c,ls}}} \right)
\Sigma_{\mathrm{c,lp}} = \left(
\frac{\Sigma_{\mathrm{c,lp}}}{\Sigma_{\mathrm{c,ls}}} \right) \Delta
\Sigma^{\mathrm{true}} .
\end{equation}
Combining many source-lens pairs allowing for a pair weight
$w_{\mathrm{ls}}$ we find,
\begin{equation}
  \hds = \frac{\sum\limits_{\mathrm{ls}} w_{\mathrm{ls}} \left(
    \frac{\Sigma_{\mathrm{c,lp}}}{\Sigma_{\mathrm{c,ls}}} \right)
    \Delta \Sigma^{\mathrm{true}}}{\sum\limits_{\mathrm{ls}} w_{\mathrm{ls}}}
  .
\end{equation}
Using the optimal weight $w_{\mathrm{ls}} \propto
\Sigma^{-2}_{\mathrm{c,lp}}$ this expression may be written,
\begin{equation}
  \hds = \frac{\sum\limits_{\mathrm{ls}} \Sigma^{-1}_{\mathrm{c,lp}} \,
    \Sigma^{-1}_{\mathrm{c,ls}} \, \Delta
    \Sigma^{\mathrm{true}}}{\sum\limits_{\mathrm{ls}}
    \Sigma^{-2}_{\mathrm{c,lp}}} .
\end{equation}
The estimated value of $\Delta \Sigma$ hence contains a multiplicative
bias, $\Delta \Sigma^{\mathrm{true}} = f_{\mathrm{bias}} \, \langle
\hds \rangle$ where,
\begin{equation}
  f_{\mathrm{bias}} = \frac{\sum\limits_{\mathrm{ls}}
    \Sigma^{-2}_{\mathrm{c,lp}}}{\sum\limits_{\mathrm{ls}}
    \Sigma^{-1}_{\mathrm{c,lp}} \, \Sigma^{-1}_{\mathrm{c,ls}}} =
  \frac{\sum\limits_{\mathrm{ls}} w_{\mathrm{ls}}}{\sum\limits_{\mathrm{ls}}
    w_{\mathrm{ls}} \, \Sigma_{\mathrm{c,lp}} \,
    \Sigma^{-1}_{\mathrm{c,ls}}} .
\label{eqfbias}
\end{equation}
This multiplicative correction factor may be estimated at each lens
redshift from a representative subset of sources with complete
spectroscopic and photometric redshift information, by evaluating the
sums in the numerator and denominator of Eq.\ \ref{eqfbias}
\citep{Nakajima12}.

An alternative formulation of the photo-$z$ dilution correction may be
derived from the statistical distance distribution of the sources.
Provided that the lens distribution is sufficiently narrow,
Eq.\ \ref{eqgtdsigbroad} indicates that an unbiased estimate of
$\Delta\Sigma$ from each lens-source pair is,
\begin{equation}
  \hds(R) = e_{\mathrm{t}}(R/\chi_{\mathrm{l}}) \, \left[
    \overline{\Sigma_{\mathrm{c}}^{-1}}(\chi_{\mathrm{l}})
    \right]^{-1} ,
\end{equation}
where $\overline{\Sigma_{\mathrm{c}}^{-1}}$ is evaluated from
Eq.\ \ref{eqavesigc} using the source distribution
$p_{\mathrm{s}}(\chi)$.  This motivates an alternative estimator
mirroring Eq.\ \ref{eqdsigest2} \citep{Sheldon04,Miyatake15,Blake16a},
\begin{equation}
\begin{split}
  \hds(R) = \frac{\sum\limits_{\mathrm{ls}} w_{\mathrm{l}} \, w_{\mathrm{s}}
    \, e_{\mathrm{t,ls}}(R/\chi_{\mathrm{l}}) \,
    \overline{\Sigma^{-1}_{\mathrm{c,ls}}}}{\sum\limits_{\mathrm{rs}}
    w_{\mathrm{r}} \, w_{\mathrm{s}} \, \left(
    \overline{\Sigma^{-1}_{\mathrm{c,rs}}} \right)^2} \\ -
  \frac{\sum\limits_{\mathrm{rs}} w_{\mathrm{r}} \, w_{\mathrm{s}} \,
    e_{\mathrm{t,rs}}(R/\chi_{\mathrm{r}}) \,
    \overline{\Sigma^{-1}_{\mathrm{c,rs}}}}{\sum\limits_{\mathrm{rs}}
    w_{\mathrm{r}} \, w_{\mathrm{s}} \, \left(
    \overline{\Sigma^{-1}_{\mathrm{c,rs}}} \right)^2} .
\end{split}
\label{eqdsigest3}
\end{equation}
The accuracy of these potential photo-$z$ dilution corrections must be
assessed via simulations, which we consider in
Sect.\ \ref{secmocktests}.  We trialled both point-based and
distribution-based correction methods in our analysis.

\section{Amplitude-ratio test}
\label{secamp}

In this section we construct test statistics which utilise the
relative amplitudes of galaxy clustering and galaxy-galaxy lensing to
test cosmological models.  We first define the input statistics for
these tests.

\subsection{Projected clustering $w_{\mathrm{p}}(R)$}

The amplitude of galaxy-galaxy lensing is sensitive to the
distribution of matter around lens galaxies, projected along the
line-of-sight.  We can obtain an analogous projected quantity for lens
galaxy clustering by integrating the 3D galaxy auto-correlation
function, $\xi_{\mathrm{gg}}$ along the line-of-sight,
\begin{equation}
  w_{\mathrm{p}}(R) = \int_{-\infty}^\infty d\Pi \,
  \xi_{\mathrm{gg}}(R,\Pi) ,
\end{equation}
where $\Pi$ is the line-of-sight separation.  This formulation has the
additional feature of reducing sensitivity of the clustering
statistics to redshift-space distortions, which modulate the apparent
radial separations $\Pi$ between galaxy pairs.

We can estimate $w_{\mathrm{p}}(R)$ for a galaxy sample by measuring
the galaxy correlation function in $(R,\Pi)$ separation bins, and
summing over the $\Pi$ direction in the range $0 < \Pi <
\Pi_{\mathrm{max}}$:
\begin{equation}
  \hwp(R) = 2 \sum_{{\mathrm{bins}} \; i} \Delta \Pi_i \,
  \hxi_{\mathrm{gg}}(R,\Pi) .
\label{eqwpest}
\end{equation}

\subsection{The Upsilon statistics, $\Upsilon_{\mathrm{gm}}(R)$ and $\Upsilon_{\mathrm{gg}}(R)$}

Eq.\ \ref{eqdsigdef} demonstrates that the amplitude of
$\Delta\Sigma(R)$ around lens galaxies depends on the surface density
of matter across a range of smaller scales from zero to $R$, and hence
on the galaxy-matter cross-correlation coefficient at these scales.
Given that this cross-correlation is a complex function which is
difficult to model from first principles, it is beneficial to reduce
this sensitivity to small-scale information using the annular
differential surface density statistic
\citep{Reyes10,Baldauf10,Mandelbaum13},
\begin{equation}
\begin{split}
  \Upsilon_{\mathrm{gm}}(R,R_0) &= \Delta\Sigma(R) - \frac{R_0^2}{R^2}
  \Delta\Sigma(R_0) \\ &= \frac{2}{R^2} \int_{R_0}^R dR' \, R' \,
  \Sigma(R') - \Sigma(R) + \frac{R_0^2}{R^2} \Sigma(R_0) ,
\end{split}
\label{equpsgm}
\end{equation}
which is defined such that $\Upsilon_{\mathrm{gm}} = 0$ at some
small-scale limit $R = R_0$, chosen to be large enough to reduce the
main systematic effects (typically, $R_0$ is somewhat larger than the
size scale of dark matter haloes).  In this sense, the cumulative
effect from the cross-correlation function at scales $R < R_0$ is
cancelled, although it is not the case that this small-scale
suppression translates to Fourier space
\citep{Baldauf10,Asgari20b,Park20}.  In any case, the efficacy of
these statistics and choice of the $R_0$ value must be validated using
simulations, as we consider below.

The corresponding quantity suppressing the small-scale contribution to
the galaxy auto-correlations is \citep{Reyes10},
\begin{equation}
\begin{split}
  & \Upsilon_{\mathrm{gg}}(R,R_0) = \rho_{\mathrm{c}} \\ & \left[
    \frac{2}{R^2} \int_{R_0}^R dR' \, R' \, w_{\mathrm{p}}(R') -
    w_{\mathrm{p}}(R) + \frac{R_0^2}{R^2} \, w_{\mathrm{p}}(R_0)
    \right] ,
\end{split}
\label{equpsgg}
\end{equation}
where $\rho_{\mathrm{c}}$ is the critical matter density.  We note
that if $w_{\mathrm{p}}$ is defined as a step-wise function in bins
$R_i$ (with bin limits $R_{i,\mathrm{min}}$ and $R_{i,\mathrm{max}}$)
then Eq.\ \ref{equpsgg} may be written in the useful form,
\begin{equation}
\Upsilon_{\mathrm{gg}}(R,R_0) = \frac{\rho_{\mathrm{c}}}{R^2}
\sum_{i=j}^k C_i \, w_{\mathrm{p}}(R_i) ,
\label{equpsgg2}
\end{equation}
where $(k,j)$ are the bins containing $(R,R_0)$, and
\begin{equation}
C_i = \begin{cases} R_{i,\mathrm{max}}^2 & i = j
  \\ R_{i,\mathrm{max}}^2 - R_{i,\mathrm{min}}^2 & j < i < k
  \\ -R_{i,\mathrm{min}}^2 & i = k \end{cases}
\end{equation}
For convenience we chose $R_0$ to coincide with the centre of a
separation bin, such that we could use the direct measurements of
$\Delta\Sigma(R_0)$ and $w_{\mathrm{p}}(R_0)$ in Eqs.\ \ref{equpsgm}
and \ref{equpsgg} without interpolation between bins (we will show
below that our results are not sensitive to the choice of $R_0$).

\subsection{The $E_{\mathrm{G}}$ test statistic}
\label{seceg}

The relative amplitudes of weak gravitational lensing and the rate of
assembly of large-scale structure depend on the `gravitational slip'
or difference between the two space-time metric potentials.  This
signature is absent in general relativity but may be significant in
modified gravity scenarios
\citep{Uzan01,Zhang07,Jain10,Bertschinger11,Clifton12}.

\citet{Zhang07} proposed that these amplitudes might be compared by
connecting the velocity field and lensing signal generated by a given
set of matter overdensities, probed via redshift-space distortions and
galaxy-galaxy lensing, respectively.  \citet{Reyes10} implemented this
consistency test by constructing the statistic,
\begin{equation}
  E_{\mathrm{G}}(R) = \frac{1}{\beta} \,
  \frac{\Upsilon_{\mathrm{gm}}(R,R_0)}{\Upsilon_{\mathrm{gg}}(R,R_0)}
  ,
\label{eqeg}
\end{equation}
where $\beta = f/b_{\mathrm{L}}$ is the redshift-space distortion
parameter which governs the observed dependence of the strength of
galaxy clustering on the angle to the line-of-sight, in terms of the
linear growth rate of a perturbation, $f = d\ln{\delta}/d\ln{a}$.
Eq.\ \ref{eqeg} is independent of the linear galaxy bias
$b_{\mathrm{L}}$ and the amplitude of matter clustering $\sigma_8$,
given that $\beta \propto 1/b_{\mathrm{L}}$, $\Upsilon_{\mathrm{gm}}
\propto b_{\mathrm{L}} \, \sigma_8^2$ and $\Upsilon_{\mathrm{gg}}
\propto b_{\mathrm{L}}^2 \, \sigma_8^2$.  The prediction of linear
perturbation theory for general relativity in a $\Lambda$CDM Universe
is a scale-independent value $E_{\mathrm{G}}(z) =
\Omega_{\mathrm{m}}(z=0)/f(z)$, although see \citet{Leonard15} for a
detailed discussion of this approxmation.

\section{Covariance of estimators}
\label{seccov}

In this section we present analytical formulations in the Gaussian
approximation for the covariance of estimates of
$\gamma_{\mathrm{t}}(\theta)$ and $\Delta \Sigma(R)$, and model how
this covariance is modulated by the presence of a survey mask (that
is, by edge effects).  Our covariance determination hence neglects
non-Gaussian and super-sample variance components.  This is a
reasonable approximation in the context of the current analysis as
these terms are subdominant (we refer the reader to Joachimi et
al.\ (in prep.) for more details on the relative amplitude of the
different covariance terms in the context of KiDS-1000).

\subsection{Covariance of average tangential shear}
\label{seccovgt}

In Appendix \ref{seccovgtap} we derive the covariance of
$\gamma_{\mathrm{t}}$ averaged within angular bins $\theta_m$ and
$\theta_n$:
\begin{equation}
  \mathrm{Cov}[\gamma_{\mathrm{t}}^{ij}(\theta_m),\gamma_{\mathrm{t}}^{kl}(\theta_n)]
  = \frac{1}{\Omega} \int \frac{d\ell \, \ell}{2\mathrm{\pi}} \,
  \sigma^2(\ell) \, \overline{J_{2,m}}(\ell) \,
  \overline{J_{2,n}}(\ell) ,
\label{eqgtcov}
\end{equation}
where $\gamma_{\mathrm{t}}^{ij}$ denotes the average tangential shear
of source sample $j$ around lens sample $i$, $\Omega$ is the total
survey angular area in steradians, and $\overline{J_{2,n}}(\ell) =
\int_{\theta_{1,n}}^{\theta_{2,n}} \frac{2\mathrm{\pi}\theta \,
  d\theta}{\Omega_n} \, J_2(\ell \theta)$, where the integral is
between the bin limits $\theta_1$ and $\theta_2$ and $\Omega_n$ is
angular area of bin $n$ (i.e.\ the area of the annulus between the bin
limits).  The variance $\sigma^2(\ell)$ is given by the expression for
Gaussian random fields
\citep[e.g.][]{Hu04,Bernstein09,Joachimi10,Krause17,Blanchard19},
\begin{equation}
\sigma^2(\ell) = C_{\mathrm{g\kappa}}^{il}(\ell) \,
C_{\mathrm{g\kappa}}^{kj}(\ell) + \left[
  C_{\mathrm{\kappa\kappa}}^{jl}(\ell) + N_{\mathrm{\kappa\kappa}}^j
  \delta^{\mathrm{K}}_{jl} \right] \, \left[
  C_{\mathrm{gg}}^{ik}(\ell) + N_{\mathrm{gg}}^i
  \delta^{\mathrm{K}}_{ik} \right] ,
\label{eqgtvar}
\end{equation}
where $\delta^{\mathrm{K}}_{ij}$ is the Kronecker delta.  The angular
auto- and cross-power spectra appearing in Eq.\ \ref{eqgtvar} may be
evaluated using the expressions in Sect.\ \ref{sectheory}, and the
noise terms are $N_{\mathrm{\kappa\kappa}}^i =
\sigma_{\mathrm{e}}^2/\overline{n}_{\mathrm{s}}^i$ and
$N_{\mathrm{gg}}^i = 1/\overline{n}_{\mathrm{l}}^i$, where
$\sigma_{\mathrm{e}}$ is the shape noise and
$\overline{n}_{\mathrm{l}}^i$ and $\overline{n}_{\mathrm{s}}^i$ are
the angular lens and source densities of sample $i$ in units of per
steradian.

\subsection{Covariance of projected mass density}
\label{seccovdsig}

The covariance of $\Delta \Sigma$ may be deduced from the covariance
of $\gamma_{\mathrm{t}}$ using $\Delta \Sigma(R) =
\gamma_{\mathrm{t}}(\theta) / \overline{\Sigma_{\mathrm{c}}^{-1}}$,
and by scaling angular separations to projected separations at an
effective lens distance $\chi_{\mathrm{l}}$ using $\theta =
R/\chi_{\mathrm{l}}$ \citep{Singh17,Dvornik18,Shirasaki18}.  We can
map multipoles $\ell$ to the projected wavevector $k =
\ell/\chi_{\mathrm{l}}$ such that,
\begin{equation}
  \mathrm{Cov}[\Delta \Sigma^{ij}(R), \Delta \Sigma^{kl}(R')] =
  \frac{1}{\Omega} \int \frac{dk \, k}{2\mathrm{\pi}} \, \sigma^2(k)
  \, \overline{J_2}(k R) \, \overline{J_2}(k R') ,
\label{eqdsigcov}
\end{equation}
where we now express the variance in terms of projected power spectra,
\begin{equation}
  \sigma^2(k) = P^{il}_{\mathrm{g\kappa}}(k) \,
  P^{kj}_{\mathrm{g\kappa}}(k) + \left[
    P^{jl}_{\mathrm{\kappa\kappa}}(k) + N^j_{\mathrm{\kappa\kappa}}
    \delta^{\mathrm{K}}_{jl} \right] \left[ P^{ik}_{\mathrm{gg}}(k) +
    N^i_{\mathrm{gg}} \delta^{\mathrm{K}}_{ik} \right] .
\end{equation}
The power spectra are given by the following relations:
\begin{equation}
  P_{\mathrm{g\kappa}}(k) = \frac{\chi_{\mathrm{l}}^2 \,
    C_{\mathrm{g\kappa}}(k\chi_{\mathrm{l}})}{\left[\overline{\Sigma_{\mathrm{c}}^{-1}}(\chi_{\mathrm{l}})\right]^2}
  \approx \overline{\rho_{\mathrm{m}}} \int d\chi \,
  p_{\mathrm{l}}(\chi) \, P_{\mathrm{gm}}(k,\chi) ,
\end{equation}
\begin{equation}
\begin{split}
& P_{\mathrm{\kappa\kappa}}(k) = \chi_{\mathrm{l}}^2 \,
  C_{\mathrm{\kappa\kappa}}(k\chi_{\mathrm{l}}) \\ & =
  \chi_{\mathrm{l}}^2 \, \overline{\rho_{\mathrm{m}}}^2 \int d\chi \,
  \left[ \frac{\overline{\Sigma_{\mathrm{c},1}^{-1}}(\chi) \,
      \overline{\Sigma_{\mathrm{c},2}^{-1}}(\chi)}{\overline{\Sigma_{\mathrm{c},1}^{-1}}(\chi_{\mathrm{l}})
      \, \overline{\Sigma_{\mathrm{c},2}^{-1}}(\chi_{\mathrm{l}})}
    \right] \left( \frac{\chi_{\mathrm{l}}^2}{\chi^2} \right) \,
  P_{\mathrm{mm}} \left( \frac{k \, \chi_{\mathrm{l}}}{\chi},\chi
  \right) ,
\end{split}
\end{equation}
\begin{equation}
P_{\mathrm{gg}}(k) = \chi_{\mathrm{l}}^2
C_{\mathrm{gg}}(k\chi_{\mathrm{l}}) \approx \int d\chi \, p_1(\chi) \,
p_2(\chi) \, P_{\mathrm{gg}}(k,\chi) ,
\end{equation}
and the noise terms are,
\begin{equation}
N_{\mathrm{\kappa\kappa}} = \frac{\sigma_{\mathrm{e}}^2 \,
  \chi_{\mathrm{l}}^2}{\overline{n}_{\mathrm{s}} \, \left[
    \overline{\Sigma_{\mathrm{c}}^{-1}}(\chi_{\mathrm{l}}) \right]^2}
, \; \; \; N_{\mathrm{gg}} =
\frac{\chi_{\mathrm{l}}^2}{\overline{n}_{\mathrm{l}}} .
\end{equation}

\subsection{Covariance of remaining statistics}

The expression for the analytical covariance of $w_{\mathrm{p}}(R)$
may be derived as \citep[see also,][]{Singh17},
\begin{equation}
\begin{split}
  &\mathrm{Cov}[w_{\mathrm{p}}(R), w_{\mathrm{p}}(R')] = \\ &\frac{2
    L_\parallel \Pi_{\mathrm{max}}}{\Omega} \int \frac{dk \,
    k}{2\mathrm{\pi}} \, \sigma^2(k) \, J_0(k R) \, J_0(k R') ,
\end{split}
\end{equation}
where $L_\parallel$ is the total co-moving depth of the lens redshift
slice and the expression for the variance is,
\begin{equation}
  \sigma^2(k) = \left[ P_{\mathrm{gg}}(k) + N_{\mathrm{gg}} \right]^2 ,
\end{equation}
where $P_{\mathrm{gg}}(k)$ and $N_{\mathrm{gg}}$ are the 2D projected
power spectra and noise as defined in Sect.\ \ref{seccovdsig}.

We determined the analytical covariance of
$\Upsilon_{\mathrm{gm}}(R,R_0)$ from the covariance of
$\Delta\Sigma(R)$:
\begin{equation}
\begin{split}
  & \mathrm{Cov}[\Upsilon_{\mathrm{gm}}(R,R_0) ,
    \Upsilon_{\mathrm{gm}}(R',R_0)] = \mathrm{Cov}[\Delta\Sigma(R) ,
    \Delta\Sigma(R')] \\ &- \frac{R_0^2}{R'^2}
  \mathrm{Cov}[\Delta\Sigma(R) , \Delta\Sigma(R_0) ] -
  \frac{R_0^2}{R^2} \mathrm{Cov}[\Delta\Sigma(R') , \Delta\Sigma(R_0)
  ] \\ &+ \frac{R_0^4}{R^2 R'^2} \mathrm{Var}[\Delta\Sigma(R_0)] .
\end{split}
\label{equpsgmcov}
\end{equation}
For the case of $\Upsilon_{\mathrm{gg}}(R,R_0)$, we propagated the
covariance using Eq.\ \ref{equpsgg2}:
\begin{equation}
\begin{split}
&\mathrm{Cov}[\Upsilon_{\mathrm{gg}}(R,R_0) ,
    \Upsilon_{\mathrm{gg}}(R',R_0)] =
  \\ &\frac{\rho_{\mathrm{c}}^2}{R^2 \, R'^2} \sum_i \sum_j C_i \, C_j
  \, \mathrm{Cov}[w_{\mathrm{p}}(R_i) , w_{\mathrm{p}}(R_j)] .
\end{split}
\label{equpsggcov}
\end{equation}

We evaluated the covariance of the $E_{\mathrm{G}}$ statistic, where
required, by assuming small fluctuations in the variables in
Eq.\ \ref{eqeg} with respect to their mean, neglecting any
correlations between the measurements:
\begin{equation}
\begin{split}
  &\frac{\mathrm{Cov}[E_{\mathrm{G}}(R) \,
      E_{\mathrm{G}}(R')]}{E_{\mathrm{G}}(R) \, E_{\mathrm{G}}(R')} =
  \frac{\mathrm{Cov}[\Upsilon_{\mathrm{gm}}(R,R_0),\Upsilon_{\mathrm{gm}}(R',R_0)]}{\Upsilon_{\mathrm{gm}}(R,R_0)
    \, \Upsilon_{\mathrm{gm}}(R',R_0)} \\ &+
  \frac{\mathrm{Cov}[\Upsilon_{\mathrm{gg}}(R,R_0),
      \Upsilon_{\mathrm{gg}}(R',R_0)]}{\Upsilon_{\mathrm{gg}}(R,R_0)
    \, \Upsilon_{\mathrm{gg}}(R',R_0)} +
  \frac{\sigma_\beta^2}{\beta^2} ,
\end{split}
\label{eqegcov}
\end{equation}
where $\sigma_\beta$ is the error in the measurement of $\beta$.  This
neglect of correlations is an approximation, justified in the case of
our dataset by the fact that the sky area used for the galaxy
clustering measurement is substantially different to the sub-sample
used for galaxy-galaxy lensing (see Joachimi et al.\ (in prep.) for a
detailed justification of this approximation), and that the projected
lens clustering measurement ($\Upsilon_{\mathrm{gg}}$) is largely
insensitive to redshift-space distortions ($\beta$) owing to the
projection over the line-of-sight separations.  We note that in our
fiducial fitting approach, we determined the scale-independent
statistic $\langle E_G \rangle$ through direct fits to
$\Upsilon_{\mathrm{gm}}$ and $\Upsilon_{\mathrm{gg}}$ as discussed in
Sect.\ \ref{secdatatests}, without requiring the covariance of
$E_G(R)$.

\subsection{Modification of noise term}

We can replace the noise terms in Sects.\ \ref{seccovgt} and
\ref{seccovdsig} with a more accurate computation using the survey
source and lens distributions.  Neglecting the random lens term (which
is not important on the small scales for which the noise term is
significant), we find that the variance associated with the
$\gamma_{\mathrm{t}}$ estimator in Eq.\ \ref{eqgtest} is
\citep[e.g.][]{Miyatake19},
\begin{equation}
\mathrm{Var}[\gamma_{\mathrm{t}}(\theta)] = \frac{\sum\limits_{\mathrm{ls}}
  w_{\mathrm{l}}^2 \, w_{\mathrm{s}}^2 \,
  \sigma_{\mathrm{e}}^2}{\left( \sum\limits_{\mathrm{rs}} w_{\mathrm{r}} \,
  w_{\mathrm{s}} \right)^2} .
\end{equation}
Likewise, the variance associated with the $\Delta \Sigma$ estimator
in Eq.\ \ref{eqdsigest1} is,
\begin{equation}
\mathrm{Var}[\Delta\Sigma(R)] = \frac{\sum\limits_{\mathrm{ls}}
  w_{\mathrm{l}}^2 \, w_{\mathrm{s}}^2 \, w_{\mathrm{ls}}^2 \,
  \sigma_{\mathrm{e}}^2 \left( \Sigma_{\mathrm{c,ls}}
  \right)^2}{\left( \sum\limits_{\mathrm{rs}} w_{\mathrm{r}} \,
  w_{\mathrm{s}} \, w_{\mathrm{rs}} \right)^2} .
\end{equation}
We adopted these noise terms in our covariance model.

\subsection{Modification for survey window}

Eqs.\ \ref{eqgtcov} and \ref{eqdsigcov} for the analytical covariance
are modified by the survey window function.  We can intuitively
understand the need for this modification by considering that, whilst
Fourier transforms assume periodic boundary conditions, the boundaries
of the survey restrict the number of source-lens pairs on scales that
are a significant fraction of the survey dimensions.

In Appendix \ref{seccovwinap} we derive how the covariance of a
cross-correlation function $\xi(r)$ between two Gaussian fields is
modified by the window function of the fields, $W_1(\vx)$ and
$W_2(\vx)$ \citep[see also,][]{Beutler17}.  We find,
\begin{equation}
\begin{split}
 & \mathrm{Cov}[\xi(r),\xi(s)] \approx \frac{A_3(r,s)}{A_2(r) \,
    A_2(s)} \\ & \frac{1}{2\mathrm{\pi}} \int dk \, k \, \left[
    P_{11}(k) \, P_{22}(k) + P^2_{12}(k) \right] \, J_0(kr) \, J_0(ks)
  ,
\end{split}
\end{equation}
where $P_{11}$, $P_{22}$ and $P_{12}$ are the auto- and cross-power
spectra of the fields and the pre-factors are given by,
\begin{equation}
\begin{split}
A_2(r) &= \int_{\mathrm{bin} \, r} d^3\vr \int d^2\vx \, W_1(\vx) \,
W_2(\vx+\vr) \\ A_3(r,s) &= \int_{\mathrm{bin} \, r} d^3\vr
\int_{\mathrm{bin} \, s} d^3\vs \int d^2\vx \, A_{12}(\vx,\vr) \,
A_{12}(\vx,\vs) ,
\end{split}
\end{equation}
where the integrals over $\vr$ and $\vs$ are performed within the
separation bin, and we have written $A_{12}(\vx,\vr) = W_1(\vx) \,
W_2(\vx+\vr)$.  We hence approximated the dependence of the covariance
on the survey window by replacing the survey area in
Eqs.\ \ref{eqgtcov} and \ref{eqdsigcov} by the expression $A_2(r) \,
A_2(s) / A_3(r,s)$.

We calculated the terms $A_2$ and $A_3$ using the mean and covariance
of the pair count $R_{\mathrm{s}}R_{\mathrm{l}}(r)$ between random
source and lens realisations \citep{Landy93}, which have respective
densities $\overline{n}_{\mathrm{s}}$ and $\overline{n}_{\mathrm{l}}$.
The mean pair count in a separation bin at scale $r$ (between $r_1$
and $r_2$), containing bin area $A_{\mathrm{bin}}(r) = \mathrm{\pi}
(r_2^2 - r_1^2)$, is
\begin{equation}
\langle R_{\mathrm{s}}R_{\mathrm{l}}(r) \rangle =
\overline{n}_{\mathrm{s}} \, \overline{n}_{\mathrm{l}} \,
A_{\mathrm{bin}}(r) \, A_2(r),
\end{equation}
which allows us to find $A_2(r)$, given that the other variables are
known.  The covariance of the pair count between separation bins $r$
and $s$ is,
\begin{equation}
\begin{split}
& \mathrm{Cov}[ R_{\mathrm{s}}R_{\mathrm{l}}(r) ,
    R_{\mathrm{s}}R_{\mathrm{l}}(s) ] = \\ & \overline{n}_{\mathrm{s}}
  \, \overline{n}_{\mathrm{l}} \, A_{\mathrm{bin}}(r) \left[ A_2(r) \,
    \delta^K_{rs} + \left( \overline{n}_{\mathrm{s}} +
    \overline{n}_{\mathrm{l}} \right) A_{\mathrm{bin}}(s) \, A_3(r,s)
    \right] ,
\end{split}
\end{equation}
which allows us to determine $A_3(r,s)$.  For all the source-lens
configurations and separation bins considered in this study, we find
that the area correction factor differs from 1.0 by less than $10\%$.

\subsection{Propagation of errors in multiplicative corrections}
\label{seccovprop}  

Galaxy-galaxy lensing measurements are subject to multiplicative
correction factors arising from shape measurement calibration (see
Sect.\ \ref{secdatakids}) and, in the case of $\Delta\Sigma$, owing to
photometric redshift dilution (see Sect.\ \ref{secphotoz}).  We
propagated the uncertainties in these correction factors, which are
correlated between different source and lens samples, into the
analytical covariance of the measurements.  Taking $\Delta\Sigma$ as
an example and writing a general amplitude correction factor as
$\alpha$, the relation between the corrected and analytical statistics
(denoted by the superscripts `corr' and `ana', respectively) is,
\begin{equation}
  \Delta\Sigma^{\mathrm{corr}}_{ijk} = \frac{\alpha_{ij} \,
    \Delta\Sigma^{\mathrm{ana}}_{ijk}}{\langle \alpha_{ij} \rangle} ,
\end{equation}
which is normalised such that $\langle
\Delta\Sigma^{\mathrm{corr}}_{ijk} \rangle =
\Delta\Sigma^{\mathrm{ana}}_{ijk}$, where $i$ denotes the lens sample,
$j$ the source sample and $k$ the separation bin.  We hence find,
\begin{equation}
\begin{split}
  \mathrm{Cov}[ \Delta\Sigma^{\mathrm{corr}}_{ijk} ,
    \Delta\Sigma^{\mathrm{corr}}_{lmn} ] &= \mathrm{Cov}[
    \Delta\Sigma^{\mathrm{ana}}_{ijk} ,
    \Delta\Sigma^{\mathrm{ana}}_{lmn} ] \left( 1 + C_{ij,lm} \right)
  \\ &+ \langle \Delta\Sigma^{\mathrm{ana}}_{ijk} \rangle \langle
  \Delta\Sigma^{\mathrm{ana}}_{lmn} \rangle \, C_{ij,lm} ,
\end{split}
\label{eqcovprop}
\end{equation}
where,
\begin{equation}
  C_{ij,lm} = \frac{\langle \alpha_{ij} \alpha_{lm} \rangle - \langle
    \alpha_{ij} \rangle \langle \alpha_{lm} \rangle}{\langle
    \alpha_{ij} \rangle \langle \alpha_{lm} \rangle} =
  \frac{\mathrm{Cov}[ \alpha_{ij}, \alpha_{lm}]}{\langle \alpha_{ij}
    \rangle \langle \alpha_{lm} \rangle} .
\label{eqcovprop2}
\end{equation}
We describe our specific implementation of these equations in the case
of the KiDS dataset in Sect.\ \ref{secdatatests}.

\section{Data}
\label{secdata}

\subsection{KiDS-1000}
\label{secdatakids}

The Kilo-Degree Survey is a large optical wide-field imaging survey
optimised for weak gravitational lensing analysis, performed with the
OmegaCAM camera on the VLT Survey Telescope at the European Southern
Observatory's Paranal Observatory.  The survey covers two regions of
sky each containing several hundred square degrees, KiDS-North and
KiDS-South, in four filters $(u,g,r,i)$.  The companion VISTA-VIKING
survey has provided complementary imaging in near-infrared bands
$(Z,Y,J,H,K_{\mathrm{s}})$, resulting in a deep, wide, nine-band
imaging dataset.

Our study is based on the fourth public data release of the project,
KiDS-1000 \citep{Kuijken19}, which comprises $1 \, 006$ deg$^2$ of
multi-band data, more than doubling the previously-available coverage.
We used an early-science release of the KiDS-1000 shear catalogues,
which was created using the exact pipeline version and PSF modelling
strategy implemented in \citet{Hildebrandt17} for the KiDS-450
release.  We note that these catalogues have not undergone any
rigorous assessment for the presence of cosmic shear systematics, but
they are sufficient for the galaxy-galaxy lensing science presented in
this paper, as this is less susceptible to systematic errors in the
lensing catalogues.  The raw pixel data was processed by the {\tt
  THELI} and {\tt ASTRO\_WISE} pipelines \citep{Erben13,deJong15}, and
source ellipticities were measured using {\tt lensfit}
\citep{Miller13}, assigning an optimal weight for each source, and
calibrated by a large suite of image simulations \citep{Kannawadi19}.
Photometric redshifts $z_{\mathrm{B}}$ were determined from the
nine-band imaging for each source using the Bayesian code {\tt BPZ}
\citep{Benitez00}, calibrated using spectroscopic sub-samples
\citep{Hildebrandt20}, and used to divide the sources into tomographic
bins according to the value of $z_{\mathrm{B}}$.

\subsection{BOSS}

The Baryon Oscillation Spectroscopic Survey \citep[BOSS,][]{Dawson13}
is the largest existing galaxy redshift survey, which was performed
using the Sloan Telescope between 2009 and 2014.  BOSS mapped the
distribution of $1.5$ million Luminous Red Galaxies (LRGs) and quasars
across $\sim 10 \, 000$ deg$^2$, inspiring a series of cosmological
analyses including the most accurate existing measurements of baryon
acoustic oscillations and redshift-space distortions in the galaxy
clustering pattern \citep{Alam17b}.  The final (Data Release 12)
large-scale structure catalogues are described by \citet{Reid16}; we
used the combined LOWZ and CMASS LRG samples in our
study.\footnote{The BOSS large-scale structure samples are available
  for download at the link
  \url{https://data.sdss.org/sas/dr12/boss/lss/}.}

\subsection{2dFLenS}

\begin{figure*}
\centering
\includegraphics[width=\textwidth]{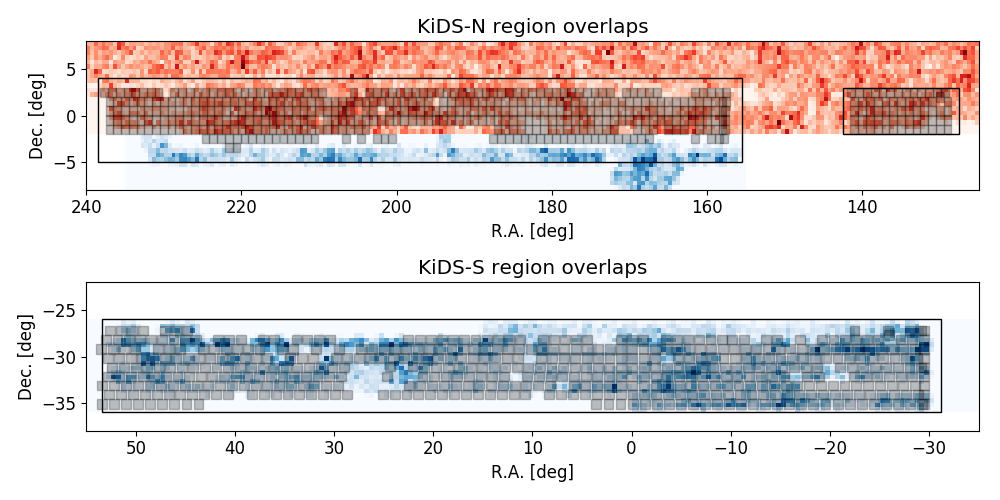}
\caption{KiDS source and LRG lens catalogues within the KiDS-N region
  (top panel) and KiDS-S region (bottom panel).  The grey squares
  represent the KiDS-1000 pointings, and the fluctuating background is
  the gridded number density of 2dFLenS (blue) and BOSS (red)
  galaxies.  The open rectangles outline the footprint of the full
  KiDS survey, indicating that the LRG overlap will continue to
  increase as the survey is completed.}
\label{figoverlap}
\end{figure*}

The 2-degree Field Lensing Survey \citep[2dFLenS,][]{Blake16b} is a
galaxy redshift survey performed at the Australian Astronomical
Observatory in 2014-2015 using the 2-degree Field spectroscopic
instrument, with the goal of extending spectroscopic-redshift coverage
of gravitational lensing surveys in the southern sky, particularly the
KiDS-South region.  The 2dFLenS sample covers an area of 731 deg$^2$
and includes redshifts for $40 \, 531$ LRGs in the redshift range $z <
0.9$, selected by applying BOSS-inspired colour-magnitude cuts to the
VST-ATLAS imaging data.\footnote{The 2dFLenS dataset is publicly
  available at the link \url{http://2dflens.swin.edu.au}.}  The
2dFLenS dataset has already been utilised in conjunction with the
KiDS-450 lensing catalogues to perform a previous implementation of
the amplitude-ratio test \citep{Amon18}, a combined cosmological
analysis of cosmic shear tomography, galaxy-galaxy lensing and galaxy
multipole power spectra \citep{Joudaki18} and to determine photometric
redshift calibration by cross-correlation
\citep{Johnson17,Hildebrandt20}.  In our study we utilised the 2dFLenS
LRG sample which overlapped with the KiDS-1000 pointings in the
southern region.  Fig.\ \ref{figoverlap} illustrates the overlaps of
the KiDS-1000 source catalogues in the north and south survey regions
with the BOSS and 2dFLenS LRG catalogues.

\section{Mocks}
\label{secmocks}

We used the {\tt MICECATv2.0} simulation
\citep{Fosalba15a,Crocce15,Fosalba15b} to produce representative KiDS
lensing source catalogues and LRG lens catalogues for testing the
estimators, models and covariances described above.  The Marenostrum
Institut de Ciencias de l'Espai (MICE) catalogues cover an octant of
the sky ($0 < \mathrm{RA} < 90^\circ$, $0 < \mathrm{Dec} < 90^\circ$)
for redshift range $z < 1.4$.  We used boundaries at constant RA and
Dec to divide this area into 10 sub-samples, each of area 516 deg$^2$.
The fiducial set of cosmological parameters for the mock is
$\Omega_{\mathrm{m}} = 0.25$, $h = 0.7$, $\Omega_{\mathrm{b}} =
0.044$, $\sigma_8 = 0.8$ and $n_{\mathrm{s}} = 0.95$.

\subsection{Mock source catalogue}

We constructed the representative mock source catalogue by applying
the following steps (see van den Busch et al.\ (in prep.) for a full
description of the MICE KiDS source mocks).  The MICE catalogue is
non-uniform across the octant: the region $\mathrm{Dec} < 30^\circ$
AND [($\mathrm{RA} < 30^\circ$) OR ($\mathrm{RA} > 60^\circ$)] has a
shallower redshift distribution than the remainder.  Firstly, we
homogenised the catalogue with the cut ${\tt
  des\_asahi\_full\_i\_true} < 24$, such that we could construct mocks
using the complete octant.  The MICE catalogue shears $(\gamma_1,
\gamma_2)$ are defined by the position angle relative to the
declination axis.  Given the MICE system for mapping 3D positions to
(RA, Dec) co-ordinates, the KiDS conventions can be recovered by the
following transformations: $\mathrm{RA} \rightarrow 90^\circ -
\mathrm{RA}$, $\gamma_1 \rightarrow -\gamma_1$ ($\gamma_2$ is
effectively negated twice and therefore unchanged).

\begin{figure*}
\centering
\includegraphics[width=\textwidth]{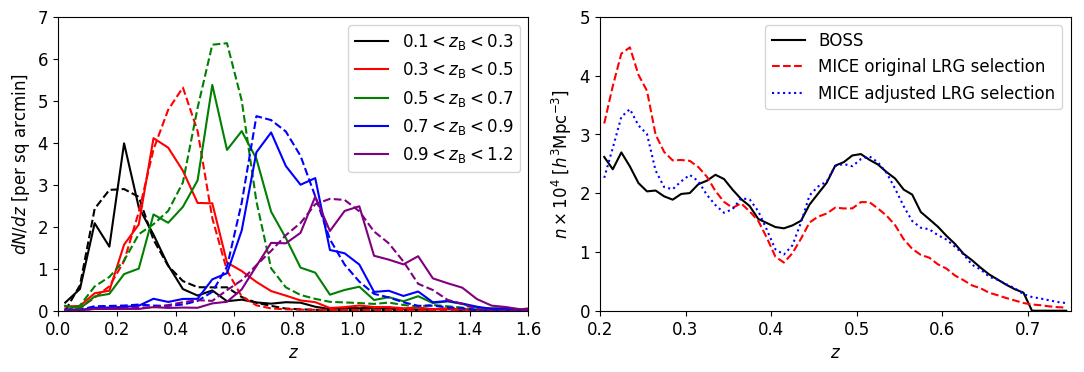}
\caption{{\it Left panel:} Redshift distribution estimates of the KiDS
  source catalogue (solid lines) and MICE mock source catalogue
  (dashed lines) in the five source tomographic bins split by
  photometric redshift.  {\it Right panel:} Number density as a
  function of redshift of the BOSS LRG dataset (black solid line), the
  MICE mock lens catalogue with the original BOSS colour selection
  (red dashed line), and the MICE mock lens catalogue with the
  adjusted BOSS colour selection (blue dotted line).}
\label{fignzmock}
\end{figure*}

We constructed a KiDS-like photometric realisation based on the galaxy
sizes and shapes, median KiDS seeing and limiting magnitudes,
including photometric noise (see van den Busch et al.\ in prep.).  We
ran BPZ photometric redshift estimation \citep{Benitez00} on the mock
source magnitudes and sizes, assigning $z_{\mathrm{B}}$ values for
each object.  We used a KDTree algorithm to assign weights to the mock
sources on the basis of a nearest-neighbour match to the data
catalogue in magnitude space, and randomly sub-sampled the catalogue
to match the KV450 effective source density.  We produced noisy shear
components $(e_1, e_2)$ as $e = (\gamma + n)/(1 + n \gamma^*)$
\citep{Seitz97} where $\gamma = \gamma_1 + \gamma_2 \, \mathrm{i}$, $e
= e_1 + e_2 \, \mathrm{i}$ and $n = n_1 + n_2 \, \mathrm{i}$, where
$n_1$ and $n_2$ are drawn from Gaussian distributions with standard
deviation $\sigma_{\mathrm{e}} = 0.288$ \citep{Hildebrandt20}.  The
redshift distribution estimates of the KiDS data and MICE mock source
tomographic samples are displayed in the left panel of
Fig.\ \ref{fignzmock}, illustrating the reasonable match between the
two catalogues.

\subsection{Mock lens catalogue}

We constructed the representative mock LRG lens catalogue from the
MICE simulation as follows.  We used the galaxy magnitudes {\tt
  sdss\_g\_true}, {\tt sdss\_r\_true}, {\tt sdss\_i\_true} and first
applied the MICE evolution correction to these magnitudes as a
function of redshift, $m \rightarrow m - 0.8*[\mathrm{arctan}(1.5*z) -
  0.1489]$ \citep{Crocce15}.  We then constructed the LRG lens
catalogues using the BOSS LOWZ and CMASS colour cuts in terms of the
variables,
\begin{equation}
\begin{split}
& c_\parallel = 0.7 \, (g-r) + 1.2 \, (r-i-0.18) , \\
& c_\perp = (r-i) - (g-r)/4 - 0.18 , \\
& d_\perp = (r-i) - (g-r)/8 .
\end{split}
\end{equation}
Applying the original BOSS colour-magnitude selection cuts
\citep{Eisenstein11} to the MICE mock did not reproduce the BOSS
redshift distribution (which is unsurprising, since this mock has not
been tuned to do so; the BOSS data is also selected from noisy
observed magnitudes).  Our approach to resolve this issue, following
\citet{Crocce15}, was to vary the colour and magnitude selection cuts
to minimise the deviation between the mock and data redshift
distributions.  We applied the following LOWZ selection cuts (where we
indicate our changed values in bold font, and the previous values
immediately following in square brackets):
\begin{equation}
\begin{split}
& 16.0 < r < \mathbf{20.0} [19.6] , \\
& r < \mathbf{13.35} [13.5] + c_\parallel/0.3 , \\
& |c_\perp| < 0.2 .
\end{split}
\end{equation}
We applied the following CMASS selection cuts:
\begin{equation}
\begin{split}
& 17.5 < i < \mathbf{20.06} [19.9] , \\
& r - i < 2 , \\
& d_\perp > 0.55 , \\
& i < \mathbf{19.98} [19.86] + 1.6 \, (d_\perp-0.8) .
\end{split}
\end{equation}
The resulting redshift distributions of the MICE lens mock (original,
and after adjustment of the colour selection cuts) and the BOSS data
are shown in the right panel of Fig.\ \ref{fignzmock}, illustrating
that our modified selection produced a much-improved representation of
the BOSS dataset.  The clustering amplitude of the MICE LRG mock
catalogues was consistent with a galaxy bias factor $b \approx 2$,
although did not precisely match the clustering of the BOSS dataset,
since it was not tuned to do so.  However, these representative
catalogues nonetheless allowed us to test our analysis procedures.

\section{Simulation tests}
\label{secmocktests}

In this section we analyse the representative source and lens
catalogues constructed from the MICE mocks described in Sect.\
\ref{secmocks}.  Our specific goals are to:
\begin{itemize}
  \item Test that the non-linear galaxy bias model specified in
    Sect.\ \ref{secbias} is adequate for modelling the galaxy-galaxy
    lensing and clustering statistics across the relevant scales.
  \item Test that the approaches to the photo-$z$ dilution correction
    of $\Delta\Sigma$ described in Sect.\ \ref{secphotoz} recovered
    results consistent with those obtained using source spectroscopic
    redshifts.
  \item Use the multiple mock realisations and jack-knife techniques
    to test that the covariance of the estimated statistics is
    consistent with the analytical Gaussian covariance specified in
    Sect.\ \ref{seccov}.
  \item Test that the $E_{\mathrm{G}}$ test statistics constructed
    from the mock as described in Sect.\ \ref{seceg} are consistent
    with the theoretical expectation, and determine the degree to
    which this result depends on the choice of the small-scale cut-off
    parameter $R_0$ (see Eq.\ \ref{equpsgm}).
  \item Test that, given our galaxy bias model, the galaxy-galaxy
    lensing and clustering statistics may be jointly described by a
    normalisation parameter $\sigma_8$ that is consistent with the
    mock fiducial cosmology, and use this test to assess the relative
    precision of angular and projected estimators.
\end{itemize}
Consistent with our subsequent data analysis, we divided the source
catalogues into five different tomographic samples by the value of the
BPZ photometric redshift, with divisions $z_{\mathrm{B}} = [0.1, 0.3,
  0.5, 0.7, 0.9, 1.2]$ \citep[following][]{Hildebrandt20}.  We divided
the lens catalogue into five slices of spectroscopic redshift
$z_{\mathrm{l}}$ of width $\Delta z_{\mathrm{l}} = 0.1$ in the range
$0.2 < z_{\mathrm{l}} < 0.7$.  This narrow spectroscopic slicing
minimises systematic effects due to redshift evolution across the lens
slice \citep{Leauthaud17,Singh19}.

\subsection{Measurements}
\label{secmicemeas}

\begin{figure*}
\centering
\includegraphics[width=\textwidth]{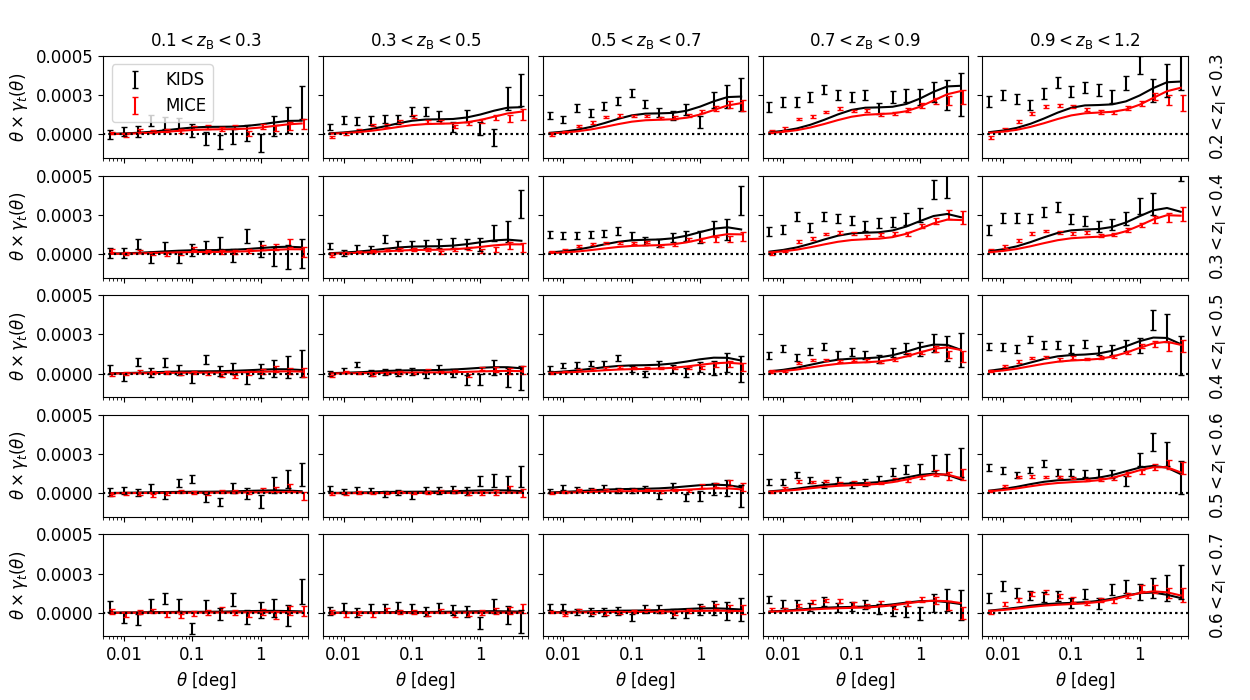}
\caption{Measurements of the average tangential shear,
  $\gamma_{\mathrm{t}}(\theta)$, for the KiDS-LRG dataset (black
  points) and representative MICE mocks (red points), between all
  pairs of lens spectroscopic redshift slices (rows) and source
  tomographic samples split by photometric redshift (columns).  The
  errors are derived from the diagonal elements of the full analytical
  covariance matrix (where we note that measurements are correlated
  across scales and samples).  The overplotted model is not a fit to
  this dataset, but rather a prediction based on the galaxy bias
  parameter fits to the $\Upsilon_{\mathrm{gg}}$ statistic of each
  lens redshift slice, which is inaccurate on small scales.  The
  $y$-values are scaled by a factor of $\theta$ (in degrees) for
  clarity.}
\label{figgttom}
\end{figure*}

We measured the following three statistics, which form the fundamental
set of measurements from which the associated statistics are derived.
Firstly, we measured the average tangential shear
$\gamma_{\mathrm{t}}(\theta)$ between all tomographic pairs of source
and lens samples, in 15 logarithmically-spaced angular bins in the
range $0.005^\circ < \theta < 5^\circ$, using the estimator of
Eq.\ \ref{eqgtest}.  This measurement is displayed in
Fig.\ \ref{figgttom} as the mean of the 10 individual mock
realisations (which each have area 516 deg$^2$), for each of the five
source samples against the five lens redshift slices.

\begin{figure*}
\centering
\includegraphics[width=\textwidth]{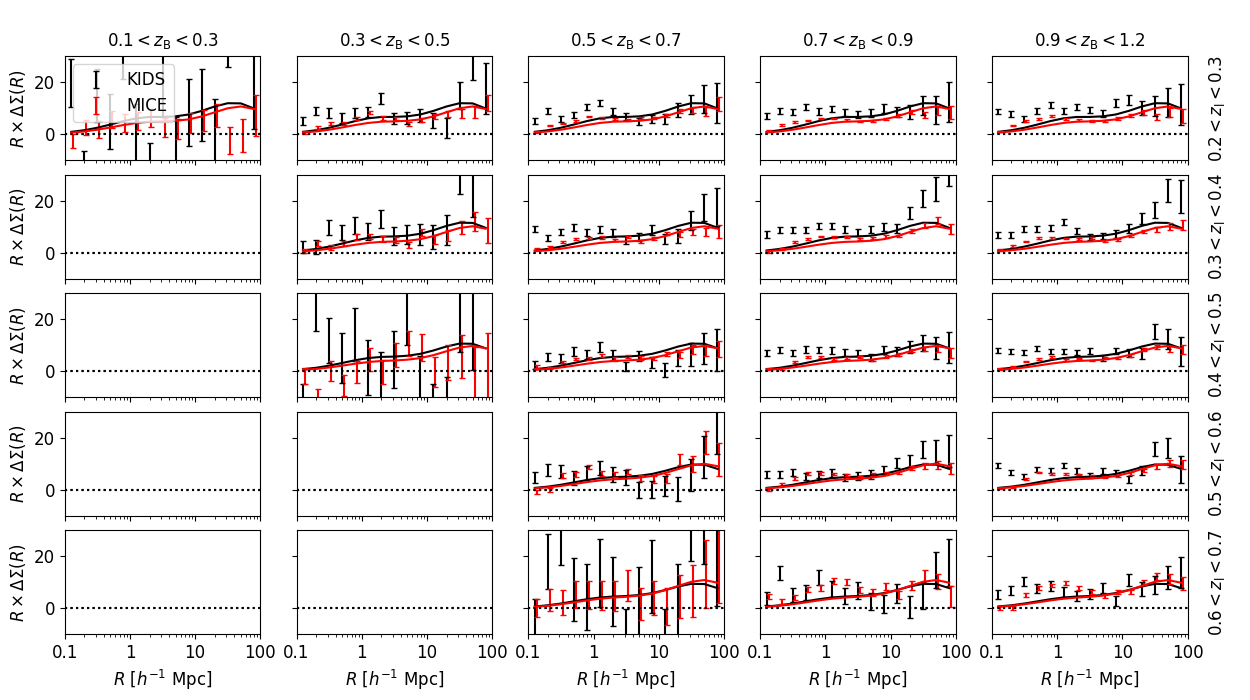}
\caption{Measurements of the projected mass density,
  $\Delta\Sigma(R)$, for the KiDS-LRG dataset (black points) and
  representative MICE mocks (red points), displayed in the same style
  as Fig.\ \ref{figgttom}.  No measurements are possible for the lower
  left-hand set of panels, owing to the adopted cut in source-lens
  pairs, $z_{\mathrm{B}} > z_{\mathrm{l}}$.  The units of
  $\Delta\Sigma$ are $h \, M_\odot$ pc$^{-2}$.}
\label{figdsigtom}
\end{figure*}

Secondly, we measured the projected mass density $\Delta\Sigma(R)$
between all tomographic pairs of source and lens samples, in 15
logarithmically-spaced projected separation bins in the range $0.1 < R
< 100 \, h^{-1}$ Mpc.  The mock mean measurement is displayed in
Fig.\ \ref{figdsigtom} in units of $h \, M_\odot$ pc$^{-2}$, for each
of the five source samples against the five lens redshift slices.
When performing a $\Delta\Sigma$ measurement between source and lens
samples we only included individual source-lens galaxy pairs with
$z_{\mathrm{B}} > z_{\mathrm{l}}$, for which the source photometric
redshift lies behind the lens spectroscopic redshift (adopting an
alternative cut $z_{\mathrm{B}} > z_{\mathrm{l}} + 0.1$ did not change
the results significantly).  We applied the photo-$z$ dilution
correction $f_{\mathrm{bias}}$ computed using Eq.\ \ref{eqfbias} based
on the point photo-$z$ values, and we study the efficacy of this
correction in Sect.\ \ref{secmicephotoz} below.

\begin{figure*}
\centering
\includegraphics[width=\textwidth]{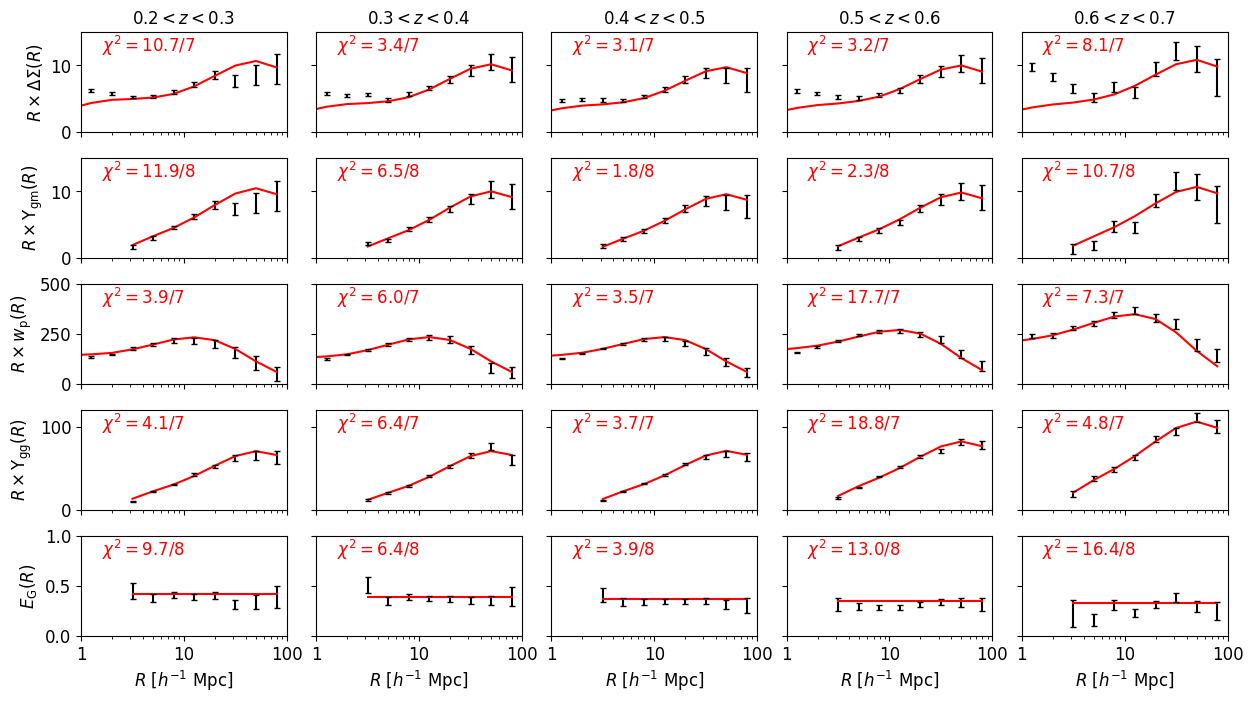}
\caption{Measurements of a series of galaxy-galaxy lensing and
  clustering statistics (rows) for each different lens redshift slice
  (columns) of the MICE mocks.  The first row displays the projected
  mass density $\Delta\Sigma(R)$, where measurements corresponding to
  the different source tomographic samples have been optimally
  combined following the procedure described in Appendix
  \ref{seccombap}.  The second row shows the corresponding
  measurements of $\Upsilon_{\mathrm{gm}}(R)$, derived from
  $\Delta\Sigma(R)$ using Eq.\ \ref{equpsgm} and assuming $R_0 = 2 \,
  h^{-1}$ Mpc.  The third and fourth rows display the galaxy
  clustering statistics $w_{\mathrm{p}}(R)$ and
  $\Upsilon_{\mathrm{gg}}(R)$ for each lens redshift slice, where
  $w_{\mathrm{p}}(R)$ is the projected correlation function
  measurement, and $\Upsilon_{\mathrm{gg}}(R)$ is the associated
  quantity suppressing contributions from small scales, derived using
  Eq.\ \ref{equpsgg} and again assuming $R_0 = 2 \, h^{-1}$ Mpc.  The
  fifth row shows the combined-probe statistic $E_{\mathrm{G}}(R)$,
  derived from the $\Upsilon_{\mathrm{gm}}(R)$ and
  $\Upsilon_{\mathrm{gg}}(R)$ measurements using Eq.\ \ref{eqeg}.
  Errors are displayed using the diagonal elements of the analytical
  covariance matrix, propagating errors where appropriate.  The
  overplotted models are determined using the galaxy bias factors
  fitted to the $\Upsilon_{\mathrm{gg}}$ measurements for each lens
  redshift slice, and $\chi^2$ statistics between the mock mean data
  and model are displayed in each panel.}
\label{figstatmice}
\end{figure*}

Thirdly, we measured the projected clustering $w_{\mathrm{p}}(R)$ of
the lens samples in the same projected separation bins as above, using
the estimator of Eq.\ \ref{eqwpest} with $\Pi_{\mathrm{max}} = 100 \,
h^{-1}$ Mpc.  The mock mean measurement of $w_{\mathrm{p}}(R)$ is
displayed as the third row of Fig.\ \ref{figstatmice}, for each of the
five redshift slices.  Each of these measurements will be compared
with cosmological model predictions as described in the subsections
below.

\begin{figure*}
\centering
\includegraphics[width=\textwidth]{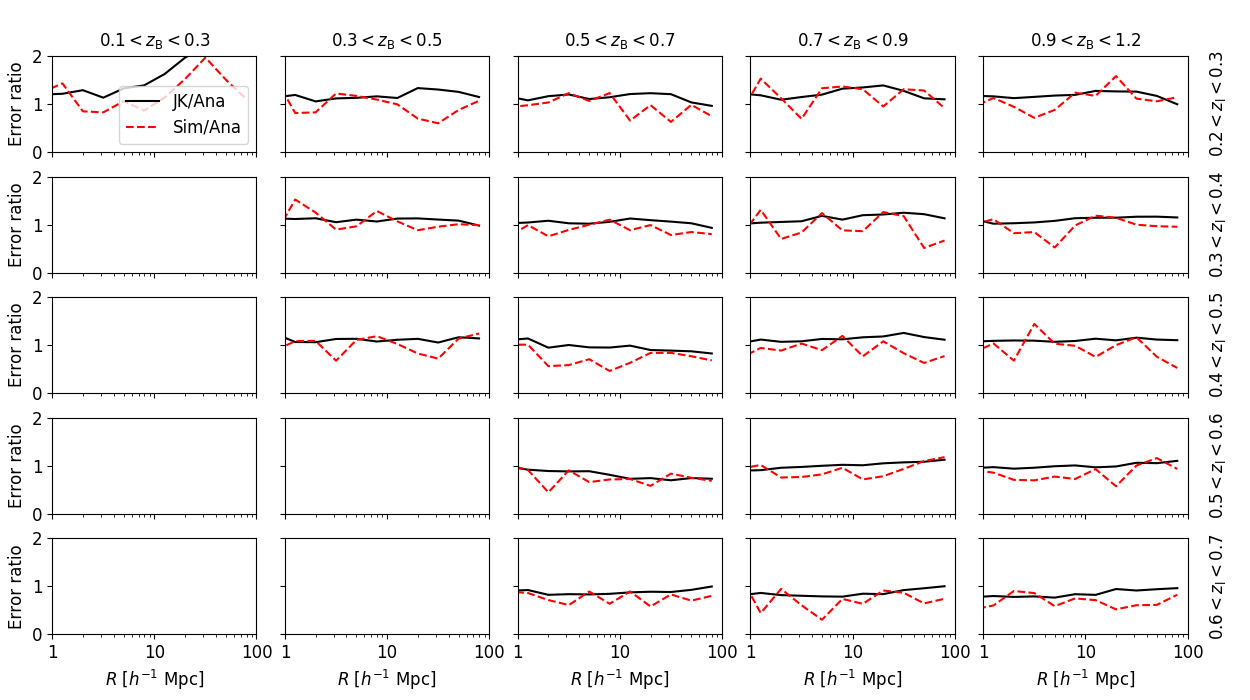}
\caption{Comparison of different estimates of the error in
  $\Delta\Sigma(R)$ for individual 516 deg$^2$ realisations of the
  MICE mocks, between all pairs of lens spectroscopic redshift slices
  (rows) and source tomographic samples (columns).  The black solid
  line shows the ratio between the error determined by a jack-knife
  analysis of the data and the error derived from the diagonal
  elements of the analytical covariance matrix, and the red dashed
  line is the ratio between the standard deviation across 10
  realisations and the analytical error.  No measurements are possible
  for the lower left-hand set of panels, owing to the adopted cut in
  source-lens pairs, $z_{\mathrm{B}} > z_{\mathrm{l}}$.}
\label{figdelsigerrmice}
\end{figure*}

We computed the covariance matrix for each statistic using the
analytical Gaussian covariance specified in Sect.\ \ref{seccov}, where
we initially used a fiducial lens linear bias factor $b_{\mathrm{L}} =
1.8$, and iterated this value following a preliminary fit to the
projected lens clustering.  Fig.\ \ref{figdelsigerrmice} compares
three different determinations of the error in $\Delta\Sigma(R)$ for
each individual 516 deg$^2$ realisation of the MICE mocks: using the
analytical covariance, using a jack-knife analysis, and derived from
the standard deviation of the 10 realisations.  For the jack-knife
analysis, we divided the sample into $7 \times 7$ angular regions
using constant boundaries in $\mathrm{RA}$ and $\mathrm{Dec}$, such
that each region contained the same angular area $10.5$ deg$^2$.  In
Fig.\ \ref{figdelsigerrmice} we display the comparison as a ratio
between the jack-knife or realisations error, and the analytical
error.

We find that in the range $R > 1 \, h^{-1}$ Mpc, where the model
provides a reasonable description of the measurements, the average
(fractional) absolute difference between the analytical and jack-knife
errors is $15\%$, and between the analytical and realisation scatter
is $21\%$ (which is the expected level of difference given the error
in the variance for 10 realisations).  Small differences between these
error estimates may arise due to the Gaussian approximation in the
analytical covariance, the exact details of the survey modelling, or
the scale of the jack-knife regions.

\begin{figure}
\resizebox{\hsize}{!}{\includegraphics{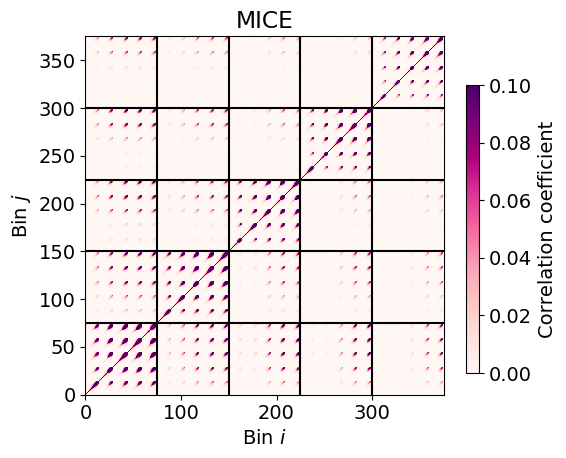}}
\caption{Full analytical covariance matrix $C_{ij}$ of the projected
  mass density $\Delta\Sigma(R)$ for the MICE mocks, spanning five
  lens redshift slices, five source tomographic samples and 15 bins of
  scale -- i.e.\ $375 \times 375$ entries -- ordered where each
  subsequent bin changes separation, then source sample, then lens
  sample, such that the vertical and horizontal solid lines demarcate
  different lens redshift slices.  We display the results as a
  correlation matrix $r = C_{ij}/\sqrt{C_{ii} \, C_{jj}}$, and note
  that there are significant off-diagonal correlations between
  measurements utilising the same lens or source sample, and between
  different scales.  We note that the colour bar is saturated at
  $r=0.1$ to reveal low-amplitude cross-correlation more clearly.}
\label{figdelsigcovfullmice}
\end{figure}

Fig.\ \ref{figdelsigcovfullmice} displays the full analytical
covariance matrix of $\Delta\Sigma(R)$ -- spanning five lens redshift
slices, five source tomographic samples and 15 bins of scale -- as a
correlation matrix with $375 \times 375$ entries.  We note that there
are significant off-diagonal correlations between measurements
utilising the same lens or source sample, and between different
scales.  The covariance matrix is reduced in dimension if source-lens
sample pairs with $z_{\mathrm{B}} < z_{\mathrm{l}}$ are excluded, as
illustrated by the missing panels in Figs.\ \ref{figdsigtom}.

We combined the correlated $\Delta\Sigma$ measurements for each
individual lens redshift slice, averaging over the five different
source tomographic samples, using the procedure described in Appendix
\ref{seccombap}.  The resulting combined $\Delta\Sigma$ measurement
for each lens redshift sample (again corresponding to a mock mean) is
shown as the first row in Fig.\ \ref{figstatmice}.

We then used the $\Delta\Sigma(R)$ and $w_{\mathrm{p}}(R)$
measurements to infer the Upsilon statistics,
$\Upsilon_{\mathrm{gm}}(R,R_0)$ and $\Upsilon_{\mathrm{gg}}(R,R_0)$,
using Eqs.\ \ref{equpsgm} and \ref{equpsgg} respectively, adopting a
fiducial value $R_0 = 2 \, h^{-1}$ Mpc (we consider the effect of
varying this choice below).  These measurements are shown in the
second and fourth rows of Fig.\ \ref{figstatmice}.  We determined the
covariance of $\Upsilon_{\mathrm{gm}}(R,R_0)$ and
$\Upsilon_{\mathrm{gg}}(R,R_0)$ using error propagation following
Eqs.\ \ref{equpsgmcov} and \ref{equpsggcov}, respectively.

Finally, we determined the $E_{\mathrm{G}}(R)$ statistic for each lens
redshift slice using Eq.\ \ref{eqeg} where, for the purposes of these
tests focussed on galaxy-galaxy lensing, we assumed a fixed input
value for the redshift-space distortion parameter $\beta =
f(z)/b_{\mathrm{L}}(z)$, where we evaluated $f(z) =
\Omega_{\mathrm{m}}(z)^{0.55}$ using the fiducial cosmology of the
MICE simulation -- we note that the exponent $0.55$ is an excellent
approximation to the solution of the differential growth equation in
$\Lambda$CDM cosmologies \citep{Linder05} -- and $b_{\mathrm{L}}(z)$
is the best-fitting linear bias parameter to the
$\Upsilon_{\mathrm{gg}}$ measurements for each lens redshift slice
$z$.  Hence, systematic errors associated with redshift-space
distortions lie beyond the scope of this study, and in our subsequent
data analysis we will infer the required $\beta$ values from existing
literature.  We propagated errors in $E_{\mathrm{G}}$ using
Eq.\ \ref{eqegcov} (and assuming no error in $\beta$ in the case of
the mocks).  Our $E_{\mathrm{G}}$ measurements are shown as the fifth
row in Fig.\ \ref{figstatmice}.

We generated fiducial cosmological models for these statistics using a
non-linear matter power spectrum $P(k,z)$ corresponding to the
fiducial cosmological parameters of the MICE simulation listed in
Sect.\ \ref{secmocks}.  We determined the best-fitting linear and
non-linear galaxy bias parameters $(b_{\mathrm{L}}, b_{\mathrm{NL}})$
by fitting to the $\Upsilon_{\mathrm{gg}}$ measurements for each lens
redshift slice for scales $R > 5 \, h^{-1}$ Mpc, and applied these
same bias parameters to the galaxy-galaxy lensing models.  The models
plotted in Figs.\ \ref{figgttom}, \ref{figdsigtom} and
\ref{figstatmice} do not otherwise contain any free parameters.  In
Fig.\ \ref{figstatmice} we display corresponding $\chi^2$ statistics
between the models and mock mean data, demonstrating a satisfactory
goodness-of-fit in general.  We evaluated the $\chi^2$ statistics for
$R > 5 \, h^{-1}$ Mpc for $\Delta\Sigma$, $w_{\mathrm{p}}$ and
$\Upsilon_{\mathrm{gg}}$, and using all scales for
$\Upsilon_{\mathrm{gm}}$ and $E_{\mathrm{G}}$.  We conclude that our
lensing and clustering measurements from the MICE mocks generally
agree with the underlying $\Lambda$CDM cosmology, which we further
explore via cosmological parameter fitting in
Sect.\ \ref{secmicecosmo}.

\subsection{Photo-$z$ dilution correction}
\label{secmicephotoz}

Within our mock analysis we considered three different implementations
of the photo-$z$ dilution correction necessary for the
$\Delta\Sigma(R)$ measurements, as described in
Sect.\ \ref{secphotoz}.  Firstly, we used the source spectroscopic
redshift values (which are available given that this is a simulation)
to produce a baseline $\Delta\Sigma$ measurement free of photo-$z$
dilution.  Secondly, for our fiducial analysis choice, we used the
source photometric redshift point values in the estimator of
Eq.\ \ref{eqdsigest2}, adopting a source-lens pair cut $z_{\mathrm{B}}
> z_{\mathrm{l}}$ and correcting for the photo-$z$ dilution using the
$f_{\mathrm{bias}}$ factor of Eq.\ \ref{eqfbias}.  We also considered
the same case, excluding the $f_{\mathrm{bias}}$ correction factor.
Thirdly, we used the redshift probability distributions for each
source tomographic sample to determine
$\overline{\Sigma_{\mathrm{c}}^{-1}}$ relative to each lens redshift
using Eq.\ \ref{eqavesigc}, and then estimated $\Delta\Sigma$ using
Eq.\ \ref{eqdsigest3}.  We refer to this as the $P(z)$
distribution-based method.

\begin{figure*}
\centering
\includegraphics[width=\textwidth]{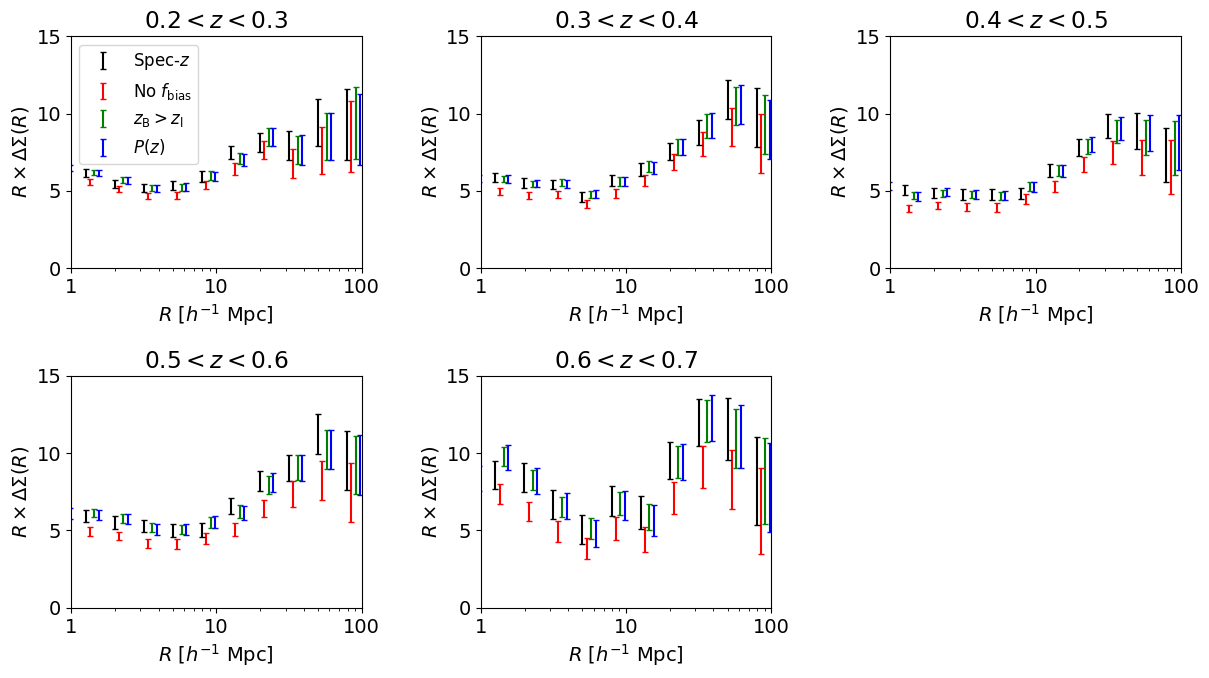}
\caption{Measurements of the projected mass density $\Delta\Sigma$ for
  each lens redshift slice of the MICE mocks, where measurements
  corresponding to the different source tomographic samples have been
  optimally combined following the procedure described in Appendix
  \ref{seccombap}.  Results are shown for four cases: using the
  spectroscopic redshifts of the sources (black points, fiducial
  measurement with perfect source redshifts), using the photometric
  redshifts of the sources but without a correction for the photo-$z$
  dilution factor $f_{\mathrm{bias}}$ (red points, photo-$z$ dilution
  remains uncorrected), including the dilution correction (green
  points), and using the redshift probability distribution for each
  source tomographic slice (blue points).  We only include individual
  source-lens pairs with $z_{\mathrm{B}} > z_{\mathrm{l}}$ in the
  measurement.  We find that, other than for the case where the
  $f_{\mathrm{bias}}$ correction is excluded, both the point-based and
  distribution-based photo-$z$ dilution corrections produce
  $\Delta\Sigma$ measurements which are consistent with those obtained
  using the source spectroscopic redshifts.}
\label{figmicephotoz}
\end{figure*}

The results of these $\Delta\Sigma$ analyses are compared in
Fig.\ \ref{figmicephotoz} for each lens redshift slice, where
measurements corresponding to the different source tomographic samples
have been optimally combined.  We find that, other than in the case
where the $f_{\mathrm{bias}}$ correction is excluded, both the
point-based and distribution-based photo-$z$ dilution corrections
produce $\Delta\Sigma$ measurements which are statistically consistent
with the baseline measurements using the source spectroscopic
redshifts.  We further verify in Sect.\ \ref{secmicecosmo} that these
analysis choices do not create significant differences in cosmological
parameter fits.

\subsection{Recovery of cosmological parameters}
\label{secmicecosmo}

Finally, we verified that our analysis methodology recovered the
fiducial cosmological parameters of the MICE simulation within an
acceptable statistical accuracy.  In this study we focus only on the
amplitudes of the clustering and lensing statistics, keeping all other
cosmological parameters fixed.  In particular we test the recovery of
the $E_{\mathrm{G}}$ statistics, and the recovery of the $\sigma_8$
normalisation, marginalising over galaxy bias parameters.

First, we determined a scale-independent $E_{\mathrm{G}}$ value (which
we denote $\langle E_{\mathrm{G}} \rangle$) for each lens redshift
slice from the MICE mock mean statistics displayed in
Fig.\ \ref{figstatmice}.  We considered two approaches to this
determination.  In one approach, we fitted a constant value to the
$E_{\mathrm{G}}(R)$ measurements shown in the fifth row of
Fig.\ \ref{figstatmice} using the corresponding analytical covariance
matrix, that is, varying a vector of five parameters,
\begin{equation}
  \vec{p} = \left[ E_G(z_1) , E_G(z_2) , E_G(z_3) , E_G(z_4) ,
    E_G(z_5) \right] .
\end{equation}
This approach has the disadvantage that it is based on the ratio of
two noisy quantities $\Upsilon_{\mathrm{gm}}/\Upsilon_{\mathrm{gg}}$,
which may result in a biased or non-Gaussian result.

Our second approach avoided this issue by including $\langle
E_{\mathrm{G}} \rangle$ as an additional parameter in a joint fit to
the $\Upsilon_{\mathrm{gm}}$ and $\Upsilon_{\mathrm{gg}}$ statistics
for each lens redshift slice, where $\langle E_{\mathrm{G}} \rangle$
changed the amplitude of $\Upsilon_{\mathrm{gm}}$ relative to
$\Upsilon_{\mathrm{gg}}$.  Specifically, we fitted the model,
\begin{equation}
\begin{split}
\Upsilon_{\mathrm{gm}}(R) &= A_{\mathrm{E}} \, b_{\mathrm{L}} \,
\Upsilon_{\mathrm{gm}}(R,\sigma_8=0.8,b_{\mathrm{L}},b_{\mathrm{NL}})
\\ \Upsilon_{\mathrm{gg}}(R) &=
\Upsilon_{\mathrm{gg}}(R,\sigma_8=0.8,b_{\mathrm{L}},b_{\mathrm{NL}})
,
\end{split}
\label{eqaefit}
\end{equation}
in terms of an amplitude parameter $A_{\mathrm{E}}$ and galaxy bias
parameters $b_{\mathrm{L}}$ and $b_{\mathrm{NL}}$, and then determined
$\langle E_{\mathrm{G}} \rangle = A_{\mathrm{E}} \, b_{\mathrm{L}} \,
E_{G,\mathrm{fid}}$ for each lens redshift slice, where
$E_{G,\mathrm{fid}}(z) = \Omega_{\mathrm{m}}/f(z)$ in terms of the
fiducial matter density parameter of the MICE mocks,
$\Omega_{\mathrm{m}}$, and the theoretical growth rate of structure
based on this matter density, $f(z)$.  Hence we vary a vector of 15
parameters,
\begin{equation}
  \vec{p} = \left[ A_{\mathrm{E},1}, b_{\mathrm{L,1}},
    b_{\mathrm{NL,1}}, ..., A_{\mathrm{E},5}, b_{\mathrm{L,5}},
    b_{\mathrm{NL,5}} \right] .
\end{equation}
We note that the $b_{\mathrm{L}}$ factor in Eq.\ \ref{eqaefit} for
$\Upsilon_{\mathrm{gm}}$ arises as a consequence of our treatment of
$\beta$ as a fixed input parameter as described in
Sect.\ \ref{secmicemeas}, and ensures that $A_{\mathrm{E}}$ is
constrained only by the relative ratio
$\Upsilon_{\mathrm{gm}}/\Upsilon_{\mathrm{gg}}$, and not the absolute
amplitude of these functions.

\begin{figure*}
\centering
\includegraphics[width=\textwidth]{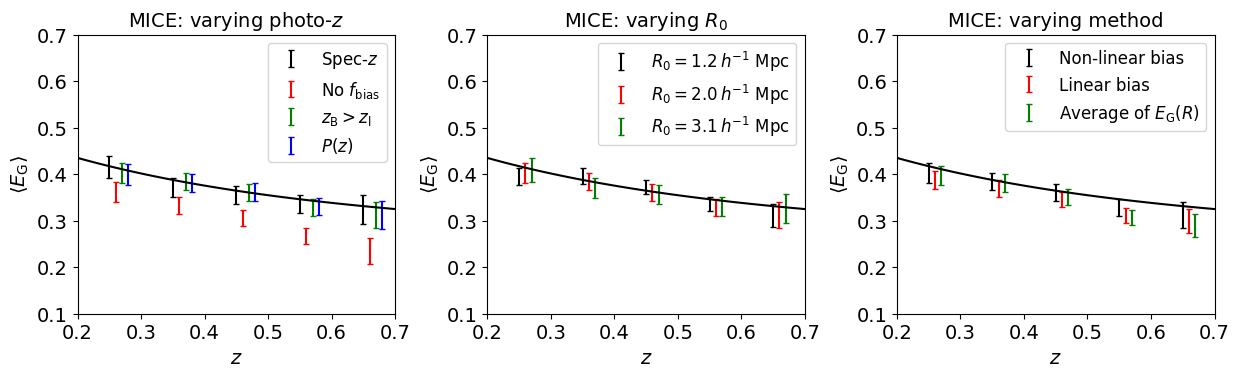}
\caption{Comparison of the scale-independent values $\langle
  E_{\mathrm{G}} \rangle$ determined for each lens redshift slice of
  the MICE mocks, varying the galaxy-galaxy lensing analysis
  assumptions and methodology.  Our fiducial analysis adopted a
  point-based photo-$z$ correction with $z_{\mathrm{B}} >
  z_{\mathrm{l}}$, $R_0 = 2.0 \, h^{-1}$ Mpc and a model fitted to
  $(\Upsilon_{\mathrm{gm}}, \Upsilon_{\mathrm{gg}})$ including
  non-linear galaxy bias.  The left panel compares determinations of
  $\langle E_{\mathrm{G}} \rangle$ varying the photo-$z$ dilution
  correction method studying the same four cases described in
  Fig.\ \ref{figmicephotoz}, the middle panel varies the small-scale
  parameter $R_0$, and the right panel alters the fitting method to
  only include linear galaxy bias, and to use a direct fit to the
  $E_{\mathrm{G}}(R)$ values.  The model line in each case is the
  prediction $E_{\mathrm{G}}(z) = \Omega_{\mathrm{m}}/f(z)$, where
  $\Omega_{\mathrm{m}}$ is the fiducial matter density parameter for
  the MICE mocks.}
\label{figmiceegvsz}
\end{figure*}

Fig.\ \ref{figmiceegvsz} displays the different determinations of
$\langle E_{\mathrm{G}} \rangle$ in each lens redshift slice.  The
left panel compares measurements using the four different treatments
of photo-$z$ dilution shown in Fig.\ \ref{figmicephotoz}, confirming
that these methods produce consistent $E_{\mathrm{G}}$ determinations
(other than the case in which $f_{\mathrm{bias}}$ is excluded; our
fiducial choice is the direct photo-$z$ pair counts with
$z_{\mathrm{B}} > z_{\mathrm{l}}$).  The middle panel compares
$\langle E_{\mathrm{G}} \rangle$ fits varying the small-scale
parameter $R_0$ (where our fiducial choice is $R_0 = 2.0 \, h^{-1}$
Mpc, and we also considered choices corresponding to the adjacent
separation bins $1.2$ and $3.1 \, h^{-1}$ Mpc).  The right panel
alters the method used to determine $\langle E_{\mathrm{G}} \rangle$,
comparing the default choice using the non-linear bias model, a linear
model where we fix $b_{\mathrm{NL}} = 0$, and a direct fit to the
scale-dependent $E_{\mathrm{G}}(R)$ values.  Reassuringly, all these
methods yielded very similar results.

We compared these determinations to the model prediction
$E_{\mathrm{G}}(z) = \Omega_{\mathrm{m}}/f(z)$ shown in
Fig.\ \ref{figmiceegvsz}.  Other than for the case where the
$f_{\mathrm{bias}}$ correction is excluded, both the point-based and
distribution-based photo-$z$ dilution corrections produce
determinations of $\langle E_{\mathrm{G}} \rangle$ which recover the
fiducial value.  This conclusion holds independently of the chosen
value of $R_0$, although higher $R_0$ values produce slightly
increased error ranges.  The different modelling approaches also
produce consistent results.

Next, we utilised our mock dataset to perform a fit of the
cosmological parameter $\sigma_8$ to the joint lensing and clustering
statistics, marginalising over different bias parameters
$(b_{\mathrm{L}}, b_{\mathrm{NL}})$ for each redshift slice such that
we vary a vector of 11 parameters,
\begin{equation}
  \vec{p} = \left[ \sigma_8, b_{\mathrm{L},1}, b_{\mathrm{NL},1}, ...,
    b_{\mathrm{L},5}, b_{\mathrm{NL},5} \right] .
\end{equation}
We fixed the remaining cosmological parameters, and performed our
parameter fit using a Markov Chain Monte Carlo method implemented
using the {\tt emcee} package \citep{ForemanMackey13}.  We used wide,
uniform priors for each fitted parameter.

As above, we adopted for our fiducial analysis the point photo-$z$
dilution correction using $f_{\mathrm{bias}}$, and we performed fits
to the $\Upsilon_{\mathrm{gm}}(R)$ and $\Upsilon_{\mathrm{gg}}(R)$
statistics with $R_0 = 2 \, h^{-1}$ Mpc, considering the same analysis
variations as above.  Following the scale cuts mentioned above, the
data vector contains eight scales for $\Upsilon_{\mathrm{gm}}(R)$ and
seven scales for $\Upsilon_{\mathrm{gg}}(R)$ for each of the five lens
redshift slices, comprising a total of 75 data points.

\begin{figure}
\resizebox{\hsize}{!}{\includegraphics{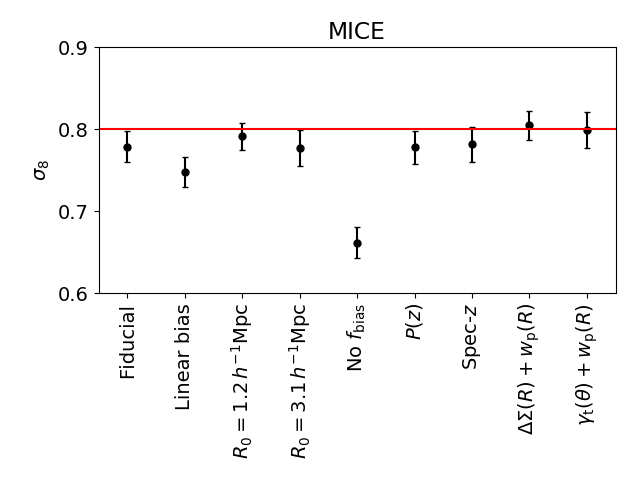}}
\caption{Measurements of the $\sigma_8$ parameter marginalising over
  the galaxy bias factors, resulting from fits to the
  $\Upsilon_{\mathrm{gm}}$ and $\Upsilon_{\mathrm{gg}}$ statistics
  measured for the MICE lens redshift slices.  We adopted a
  2-parameter galaxy bias model $(b_{\mathrm{lin}}, b_{\mathrm{NL}})$
  for each redshift slice, such that the fit varied 11 parameters in
  total.  The left-hand point resulted from our fiducial analysis
  choice, assuming $R_0 = 2.0 \, h^{-1}$ Mpc and a point-based
  photo-$z$ dilution correction.  The second point from the left
  compares the result of a fit using only the linear galaxy bias
  parameter (setting $b_{\mathrm{NL}} = 0$).  We also show results
  varying the value of $R_0$ and the photo-$z$ dilution correction.
  The horizontal red line indicates the MICE simulation fiducial
  cosmology $\sigma_8 = 0.8$.}
\label{figmicesig8}
\end{figure}

For this fiducial case, we obtained a measurement $\sigma_8 = 0.779
\pm 0.019$, consistent with the MICE simulation cosmology $\sigma_8 =
0.8$.  The $\chi^2$ statistic of the best-fitting model is $69.9$ for
64 degrees of freedom (d.o.f.), that is, 75 data points minus the 11
fitted parameters.  Fig.\ \ref{figmicesig8} displays the dependence of
the $\sigma_8$ measurements on the analysis choices.  All
methodologies using the non-linear bias model recovered the fiducial
$\sigma_8$ value, with the exception of excluding the
$f_{\mathrm{bias}}$ correction.  Adopting a linear bias model instead
produced a significantly poorer recovery.

We also considered fitting to different pairs of lensing-clustering
statistics: $\Delta\Sigma(R)$ and $w_{\mathrm{p}}(R)$ for $R >
R_{\mathrm{min}} = 5 \, h^{-1}$ Mpc, compared to
$\gamma_{\mathrm{t}}(\theta)$ and $w_{\mathrm{p}}(R)$, where we
applied a minimum-scale cut in $\theta$ which matches
$R_{\mathrm{min}}$ in each lens redshift slice.  These alternative
statistics also successfully recovered the fiducial value of
$\sigma_8$, with errors of $0.018$ (for $\Delta\Sigma$) and $0.022$
(for $\gamma_{\mathrm{t}}$).  According to this analysis, the
projected statistics produced a $\sim 20\%$ more accurate $\sigma_8$
value than the angular statistics, in agreement with the results of
\citet{Shirasaki18}.

We conclude this section by noting that the application of our
analysis pipeline to the MICE lens and source mocks successfully
recovered the fiducial $E_{\mathrm{G}}$ and $\sigma_8$ parameters of
the simulation, and is robust against differences in photo-$z$
dilution correction, choice of the small-scale parameter $R_0$, and
choice of statistic included in the analysis
[$\gamma_{\mathrm{t}}(\theta)$, $\Delta \Sigma(R)$ or
  $\Upsilon_{\mathrm{gm}}(R)$].

\section{Results}
\label{secdatatests}

\subsection{Measurements}

We now summarise the galaxy-galaxy lensing and clustering measurements
we generated from the KiDS-1000 and overlapping LRG datasets.  We cut
these catalogues to produce overlapping subsets for our galaxy-galaxy
lensing analysis, by only retaining sources and lenses within the set
of KiDS pointings which contain BOSS or 2dFLenS galaxies.  The
resulting KiDS-N sample comprised $15 \, 150 \, 250$ KiDS shapes and
$47 \, 332$ BOSS lenses within 474 KiDS pointings with total unmasked
area $366.0$ deg$^2$, and the KiDS-S sample consisted of $16 \, 994 \,
252$ KiDS shapes and $18 \, 903$ 2dFLenS lenses within 478 KiDS
pointings with total unmasked area $382.1$ deg$^2$.  We also utilised
BOSS and 2dFLenS random catalogues in our analysis, with the same
selection cuts and size 50 times bigger than the datasets, sub-sampled
from the master random catalogues provided by \citet{Reid16} and
\citet{Blake16b}, respectively.

We split the KiDS-1000 source catalogue into five different
tomographic samples by the value of the BPZ photometric redshift,
using the same bin divisions $z_{\mathrm{B}} = [0.1, 0.3, 0.5, 0.7,
  0.9, 1.2]$ adopted in Sect.\ \ref{secmocktests}.  The effective
source density of each tomographic sample is $n_{\mathrm{eff}} =
[0.88, 1.33, 2.04, 1.49, 1.26]$ arcmin$^{-2}$ \citep{Hildebrandt20},
estimated using the method of \citet{Heymans12}.  We divided the BOSS
and 2dFLenS LRG catalogues into five spectroscopic redshift slices of
width $\Delta z_{\mathrm{l}} = 0.1$ in the range $0.2 < z_{\mathrm{l}}
< 0.7$.

We measured the average tangential shear $\gamma_{\mathrm{t}}(\theta)$
and projected mass density $\Delta\Sigma(R)$ between all pairs of
KiDS-1000 tomographic source samples and LRG redshift slices in the
north and south regions, using the same estimators and binning as
utilised for the MICE mocks in Sect.\ \ref{secmicemeas} and applying a
multiplicative shear bias correction for each tomographic sample
\citep{Kannawadi19}.  For the $\Delta\Sigma$ measurement, we again
restricted the source-lens pairs such that $z_{\mathrm{B}} >
z_{\mathrm{l}}$, and (in our fiducial analysis) applied a point-based
photo-$z$ dilution correction.

We generated an analytical covariance matrix for each measurement,
initially using a fiducial lens linear bias factor $b_{\mathrm{L}} =
2$, and iterating this value following a preliminary fit to the
projected lens clustering.  We tested that the analytical error
determination agreed sufficiently well with a jack-knife error
analysis where the regions were defined as the KiDS pointings; the
results of this test and the overall analytical covariance are
visually similar to their equivalents for the MICE mocks shown in
Figs.\ \ref{figdelsigerrmice} and \ref{figdelsigcovfullmice}, and we
do not repeat these figures.  We used the KV450 spectroscopic
calibration sample with DIR weights \citep{Hildebrandt20} to estimate
the redshift distribution of each tomographic source sample for use in
the analytical covariance matrix, in the modelling of
$\gamma_{\mathrm{t}}(\theta)$ and in the distribution-based correction
to $\Delta\Sigma(R)$ for photo-$z$ dilution, and to determine the
$f_{\mathrm{bias}}$ values for the point-based photo-$z$ dilution
correction.

We propagated the uncertainties in the multiplicative correction
factors due to the shear calibration bias and photometric redshift
dilution using the method described in Sect.\ \ref{seccovprop}.
Regarding the multiplicative shear calibration, we followed
\citet{Hildebrandt20} in adopting an error $\sigma_{\mathrm{m}} =
0.02$ that is fully correlated across all samples, such that
$\mathrm{Cov}[ \alpha_{ij}, \alpha_{lm} ] = \sigma_{\mathrm{m}}^2$ in
Eq.\ \ref{eqcovprop2}.

The sample variance in the spectroscopic training set can be
characterised by an uncertainty in mean spectroscopic redshift which
varies for each tomographic sample in the range $\sigma_{\mathrm{z}} =
0.011 \rightarrow 0.039$ \citep[see][Table 2]{Hildebrandt20}.  We
propagated these errors into the determination of $f_{\mathrm{bias}}$
by re-evaluating Eq.\ \ref{eqfbias} shifting all the spectroscopic
redshifts by a small amount to determine the derivatives $\partial
f_{\mathrm{bias},ij}/\partial z_j$, where $i$ denotes the lens sample
and $j$ the source sample.  Using error propagation, we then scaled
the derivatives by the errors $\sigma_{\mathrm{z},j}$ to find the
covariance matrix of the uncertainties,
\begin{equation}
\mathrm{Cov}[ f_{\mathrm{bias},ij} \, f_{\mathrm{bias},lm} ] =
\frac{\partial f_{\mathrm{bias},ij}}{\partial z_j} \, \frac{\partial
  f_{\mathrm{bias},lm}}{\partial z_m} \, \sigma^2_{\mathrm{z},j} \,
\delta^{\mathrm{K}}_{jm} ,
\end{equation}
where the final Kronecker delta $\delta^{\mathrm{K}}_{jm}$ indicates
that these uncertainities are correlated for different lens samples
corresponding to the same source sample, but uncorrelated between
source samples (we refer the reader to Joachimi et al.\ (in prep.) for
further investigation of this point).  This uncertainty can be
propagated into the analytical covariance matrix using
Eqs.\ \ref{eqcovprop} and \ref{eqcovprop2} with $\mathrm{Cov}[
  \alpha_{ij}, \alpha_{lm} ] = \mathrm{Cov}[ f_{\mathrm{bias},ij} \,
  f_{\mathrm{bias},lm} ]$.

We used the analytical covariance matrices to combine the separate
KiDS-N and KiDS-S measurements into a single joint estimate of the
galaxy-galaxy lensing statistics and associated covariance, which we
utilised in the remainder of this study (we test the consistency of
the individual BOSS and 2dFLenS results in Sect.\ \ref{seckidssys}).
We display the KiDS $\gamma_{\mathrm{t}}(\theta)$ and
$\Delta\Sigma(R)$ galaxy-galaxy lensing measurements in the different
tomographic combinations in Figs.\ \ref{figgttom} and
\ref{figdsigtom}.  We note again that there are some differences
between the galaxy-galaxy lensing signals measured in the mocks and
data, given that the mocks have not been tuned to reproduce the BOSS
and 2dFLenS clustering properties.  These differences are particularly
evident on the smallest scales, owing to an inconsistent halo
occupation.  Our study does not require the mocks to precisely
replicate the data in order to test our analysis framework.

We obtained the most accurate measurement of the projected correlation
function $w_{\mathrm{p}}(R)$ of each lens redshift slice using the
full BOSS DR12 dataset, combining the LOWZ and CMASS selections and
spanning $9 \, 376$ deg$^2$ \citep{Reid16}.  We adopted the same
spatial separation bins as for the MICE mocks, again assuming
$\Pi_{\mathrm{max}} = 100 \, h^{-1}$ Mpc.  When analysing the BOSS
sample we included completeness weights but excluded `FKP' weights
\citep{Feldman94}, which are designed to optimise the clustering
signal-to-noise ratio but may not be appropriate in the case of
galaxy-galaxy lensing.  The full 2dFLenS dataset is too small to offer
a competitive measurement of $w_{\mathrm{p}}(R)$, although given that
it was selected using BOSS-inspired colour-magnitude cuts, we assumed
that the BOSS clustering is representative of the combined LRG sample
(and we test this approximation in Sect.\ \ref{seckidssys}).  Since
the overlap of the KiDS-N source catalogue and full BOSS sample is
small ($4\%$ of BOSS), we also assumed that the galaxy-galaxy lensing
and clustering measurements are uncorrelated.

\begin{figure*}
\centering
\includegraphics[width=\textwidth]{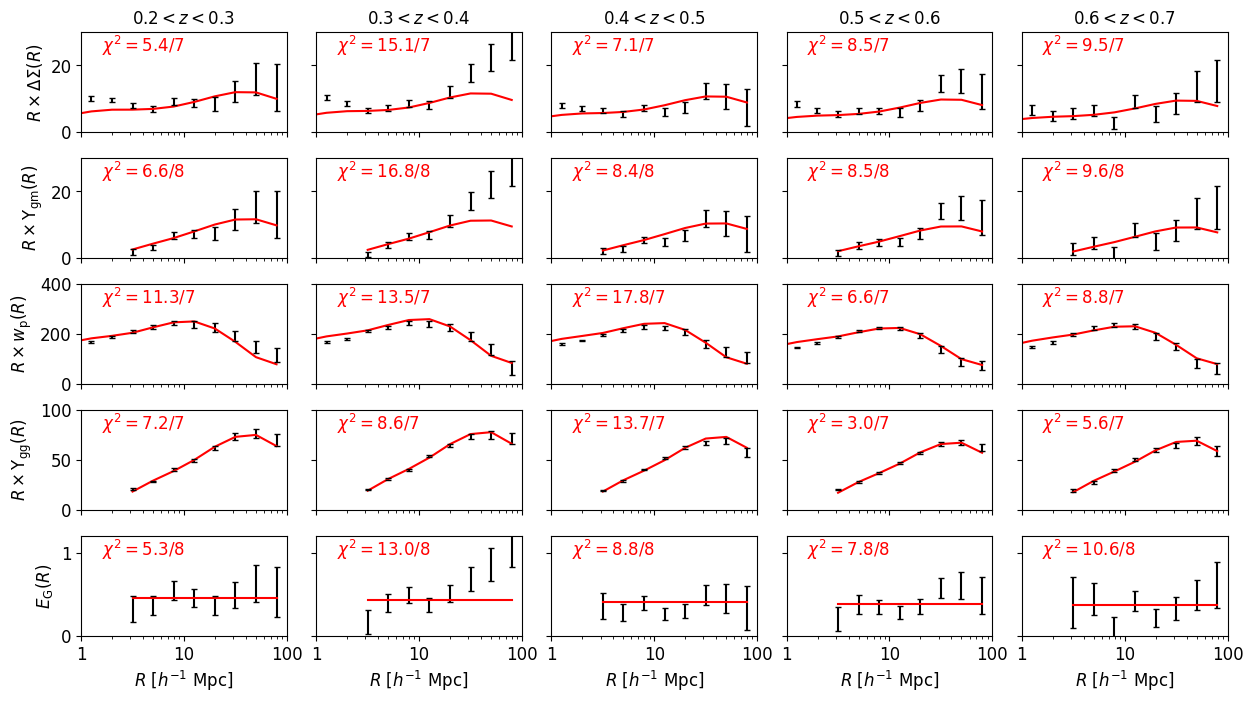}
\caption{Measurements of a series of galaxy-galaxy lensing and
  clustering statistics (rows) for each different LRG redshift slice
  (columns) correlated with the KiDS-1000 source sample, displayed in
  the same style as Fig.\ \ref{figstatmice}.  We note that the first
  and third rows -- $\Delta\Sigma(R)$ and $w_{\mathrm{p}}(R)$ --
  represent the original galaxy-galaxy lensing and projected
  clustering measurements, and the remaining rows --
  $\Upsilon_{\mathrm{gm}}(R)$, $\Upsilon_{\mathrm{gg}}(R)$ and
  $E_{\mathrm{G}}(R)$ -- represent statistics derived from these
  original measurements.}
\label{figstatkids}
\end{figure*}

We combined the correlated $\Delta\Sigma$ measurements for each lens
redshift slice, averaging over the different source samples.
Fig.\ \ref{figstatkids} displays these measurements, together with the
projected clustering $w_{\mathrm{p}}(R)$ of the full BOSS sample, the
corresponding $\Upsilon_{\mathrm{gm}}(R,R_0)$ and
$\Upsilon_{\mathrm{gg}}(R,R_0)$ statistics assuming a fiducial choice
$R_0 = 2 \, h^{-1}$ Mpc (we consider the impact of varying this choice
in Sect.\ \ref{seckidssys}), and the direct $E_{\mathrm{G}}(R)$
estimate using Eq.\ \ref{eqeg}.

We generated fiducial cosmological models for these statistics using a
non-linear matter power spectrum corresponding to the best-fitting
`TTTEEE+lowE+lensing' {\it Planck} cosmological parameters
\citep{Planck18}.  When producing the models overplotted in
Figs.\ \ref{figgttom}, \ref{figdsigtom} and \ref{figstatkids}, we
determined best-fitting linear and non-linear galaxy bias parameters
by fitting to the $\Upsilon_{\mathrm{gg}}$ measurements for each lens
redshift slice for the separation range $R > 5 \, h^{-1}$ Mpc, and
applied these same bias parameters to the galaxy-galaxy lensing
models.  Values of the $\chi^2$ statistic for each statistic and lens
redshift slice, produced using these models and the analytical
covariance, are displayed in each panel of Fig.\ \ref{figstatkids},
and indicate that the measurements are consistent with the model.

\subsection{Redshift-space distortion inputs}

We adopted values of the redshift-space distortion parameters $\beta$
for the BOSS sample as a function of redshift by interpolating the
literature analysis of \citet{Zheng19}, who provide RSD measurements
in narrow redshift slices \citep[which also agree with the compilation
  of results in][]{Alam17b}.  In order to interpolate these
measurements to our redshift locations, which slightly differ from the
bin centres of \citet{Zheng19}, we created a Gaussian process model
for $\beta(z)$ and its errors, which lie in the range $12-20\%$ as a
function of redshift, using the sum of a Matern kernel and white noise
kernel.  We note that the error in $\beta$ makes up roughly half the
variance budget for $E_{\mathrm{G}}$ in the lowest lens redshift slice
(i.e.\ increases the total error by $\sim \sqrt{2}$), but is
subdominant for the other redshift slices.

\subsection{Amplitude-ratio test $E_{\mathrm{G}}$}
\label{secegkids}

We used the KiDS-1000 and LRG clustering and galaxy-galaxy lensing
measurements, with the previously-published values of $\beta$, to
determine a scale-independent value of the amplitude ratio, $\langle
E_{\mathrm{G}} \rangle$.  We adopted the same fiducial analysis method
as for the MICE mocks: we performed a joint fit to the
$\Upsilon_{\mathrm{gm}}$ and $\Upsilon_{\mathrm{gg}}$ measurements for
each redshift slice, varying $A_{\mathrm{E}}$ and the bias parameters
$b_{\mathrm{L}}$ and $b_{\mathrm{NL}}$ as in Eq.\ \ref{eqaefit},
treating $A_{\mathrm{E}}$ as an additional amplitude parameter for
$\Upsilon_{\mathrm{gm}}$.  We then deduced $\langle E_{\mathrm{G}}
\rangle = A_{\mathrm{E}}/\beta$, propagating the errors in $\beta$
assuming all these statistics are independent (please see
Sect.\ \ref{seceg} for a note on this approximation).  We used the
analytical covariance matrices for these statistics, assumed $R_0 = 2
\, h^{-1}$ Mpc, and fitted the model to $R > 5 \, h^{-1}$ Mpc for
$\Upsilon_{\mathrm{gg}}$ and to all scales for
$\Upsilon_{\mathrm{gm}}$.  We consider the effect of varying these
analysis choices in Sect.\ \ref{seckidssys}.  In particular, we note
that fitting the directly-determined $E_{\mathrm{G}}(R)$ values (shown
in the fifth row of Fig.\ \ref{figstatkids}) produced results which
were entirely consistent with our fiducial analysis.

\begin{figure*}
\centering
\includegraphics[width=\textwidth]{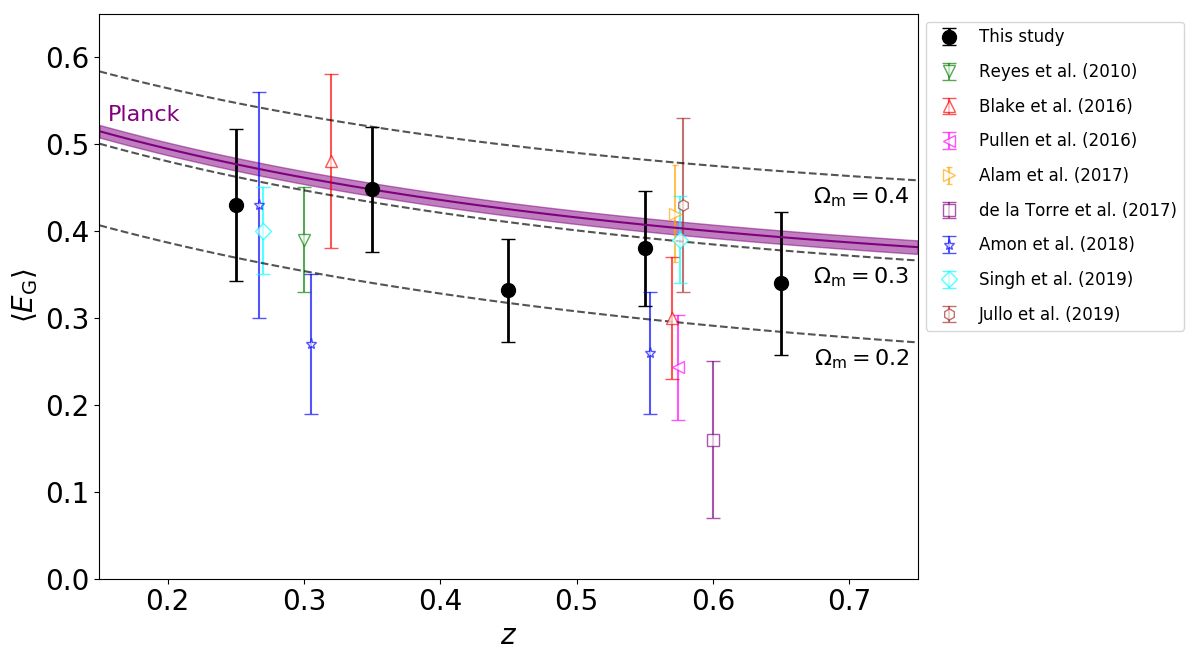}
\caption{Compilation of measurements of a scale-independent $\langle
  E_{\mathrm{G}} \rangle$ as a function of redshift.  The solid black
  data points illustrate the results of this study, in five redshift
  bins.  These are compared to literature values from: \citet{Reyes10}
  (green), \citet{Blake16a} (red), \citet{Pullen16} (magenta),
  \citet{Alam17a} (orange), \citet{delaTorre17} (purple),
  \citet{Amon18} (blue), \citet{Singh19} (cyan) and \citet{Jullo19}
  (brown).  We note that these previous measurements typically
  correspond to significantly wider lens redshift ranges than our
  study, and are also correlated in some cases owing to overlapping
  datasets.  The thickness of the purple shaded stripe illustrates the
  $68\%$ confidence range of the prediction of the {\it Planck}
  `TTTEEE+lowE+lensing' parameter chain at each redshift assuming a
  flat $\Lambda$CDM Universe, and the dashed lines are the predictions
  of the model $E_{\mathrm{G}}(z) = \Omega_{\mathrm{m}}/f(z)$ for
  $\Omega_{\mathrm{m}} = [0.2, 0.3, 0.4]$.}
\label{figegall}
\end{figure*}

\begin{figure*}
\centering
\includegraphics[width=\textwidth]{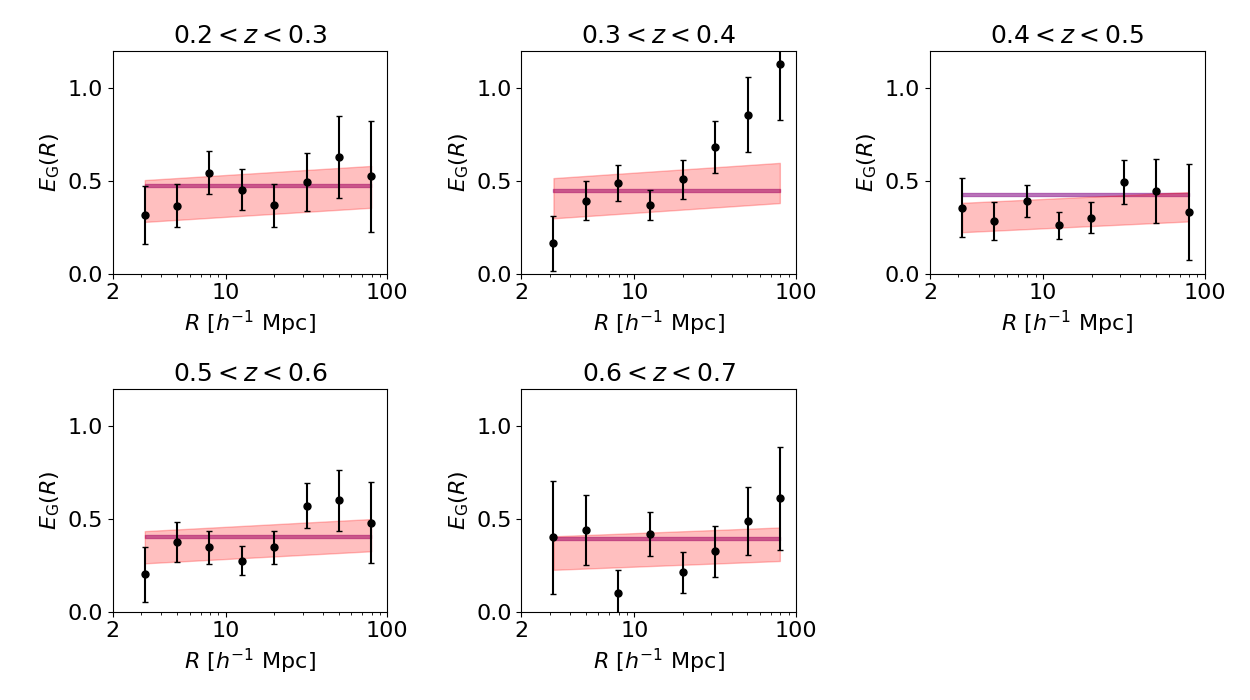}
\caption{Measurements of the $E_{\mathrm{G}}(R)$ statistic for the
  KiDS-1000 source sample in each LRG lens redshift slice.  Errors are
  derived from the diagonal elements of the analytical covariance
  matrix, propagating errors where appropriate.  The thickness of the
  purple horizontal stripes illustrates the $68\%$ confidence range of
  the prediction of the {\it Planck} `TTTEEE+lowE+lensing' parameter
  chain at each redshift assuming a flat $\Lambda$CDM Universe, and
  the red bands illustrate the $68\%$ confidence ranges of a
  scale-dependent model $E_{\mathrm{G}}(R,z_i) = A_i \left[ 1 + \alpha
    \, \log_{10}(R) \right]$ discussed in Sect.\ \ref{secegkids}.}
\label{figegscale}
\end{figure*}

Our resulting fits for $\langle E_{\mathrm{G}} \rangle$ were $[0.43
  \pm 0.09, 0.45 \pm 0.07, 0.33 \pm 0.06, 0.38 \pm 0.07, 0.34 \pm
  0.08]$ for redshifts $z = [0.25, 0.35, 0.45, 0.55, 0.65]$.  The
measurements have a small degree of correlation, owing to sharing a
common source sample, and the analytical covariance matrix is listed
in Table \ref{tabegcov}.  We plot these measurements in
Fig.\ \ref{figegall}, together with a literature compilation
\citep{Reyes10,Blake16a,Pullen16,Alam17a,delaTorre17,Amon18,Singh19,Jullo19}.
The thickness of the purple shaded stripe in Figs.\ \ref{figegall} and
\ref{figegscale} illustrates the $68\%$ confidence range of the
prediction of the {\it Planck} `TTTEEE+lowE+lensing' parameter chain
at each redshift, assuming a flat $\Lambda$CDM Universe.

\begin{table}
\caption{The covariance matrix corresponding to our measurements
  $E_{\mathrm{G}} = [0.43, 0.45, 0.33, 0.38, 0.34]$ at $z = [0.25,
    0.35, 0.45, 0.55, 0.65]$.  Each entry has been multiplied by
  $10^4$ for clarity of display.  The cross-correlation between
  different redshift slices is small.}
\label{tabegcov}
\centering
\begin{tabular}{cccccc}
\hline
Redshift & $0.25$ & $0.35$ & $0.45$ & $0.55$ & $0.65$ \\
\hline
$0.25$ & $76.821$ & $3.477$ & $3.251$ & $3.330$ & $3.178$ \\
$0.35$ & $3.477$ & $52.101$ & $1.559$ & $1.633$ & $1.509$ \\
$0.45$ & $3.251$ & $1.559$ & $34.674$ & $1.088$ & $1.017$ \\
$0.55$ & $3.330$ & $1.633$ & $1.088$ & $43.179$ & $0.893$ \\
$0.65$ & $3.178$ & $1.509$ & $1.017$ & $0.893$ & $67.585$ \\
\hline
\end{tabular}
\end{table}

Our measurements provide the best existing determination of the
lensing-clustering amplitude ratio (noting that previous measurements
displayed in Figure \ref{figegall} typically correspond to
significantly wider lens redshift ranges than our study), which is
consistent with matter density values $\Omega_{\mathrm{m}} \sim 0.3$.
Varying the $\Omega_{\mathrm{m}}$ parameter within a flat $\Lambda$CDM
cosmological model, assuming $E_{\mathrm{G}} = \Omega_{\mathrm{m}}/f =
\Omega_{\mathrm{m}}(0)/\Omega_{\mathrm{m}}(z)^{0.55}$, we find
$\Omega_{\mathrm{m}} = 0.27 \pm 0.04$ (with a minimum $\chi^2 = 1.2$
for four d.o.f.).  In principle in linear theory, this measurement is
insensitive to the other cosmological parameters in a flat
$\Lambda$CDM scenario.  The resulting error in $\Omega_{\mathrm{m}}$
is, naturally, somewhat larger than that provided by analyses
utilising the full shape of the cosmic shear and clustering functions
\citep[e.g.][]{Troester20}, albeit requiring fewer model assumptions.

We tested for the scale dependence of the $E_{\mathrm{G}}(R)$
measurements (in the fifth row of Fig.\ \ref{figstatkids}) by jointly
fitting an empirical six-parameter model $E_{\mathrm{G}}(R,z_i) = A_i
\left[ 1 + \alpha \, \log_{10}(R) \right]$ to all the redshift slices,
where $\alpha$ quantifies the fractional variation in $E_{\mathrm{G}}$
per decade in projected scale $R$ (in $h^{-1}$ Mpc) and $A_i$ is a
free amplitude for each of the five lens redshift slices.  We obtained
a $68\%$ confidence region $\alpha = 0.17 \pm 0.26$ (with a minimum
$\chi^2 = 35.9$ for 34 d.o.f.), which is consistent with no scale
dependence, as predicted in the standard gravity scenario.  These fits
are displayed in Fig.\ \ref{figegscale}.

\subsection{Systematics tests}
\label{seckidssys}

\begin{figure*}
\centering
\includegraphics[width=\textwidth]{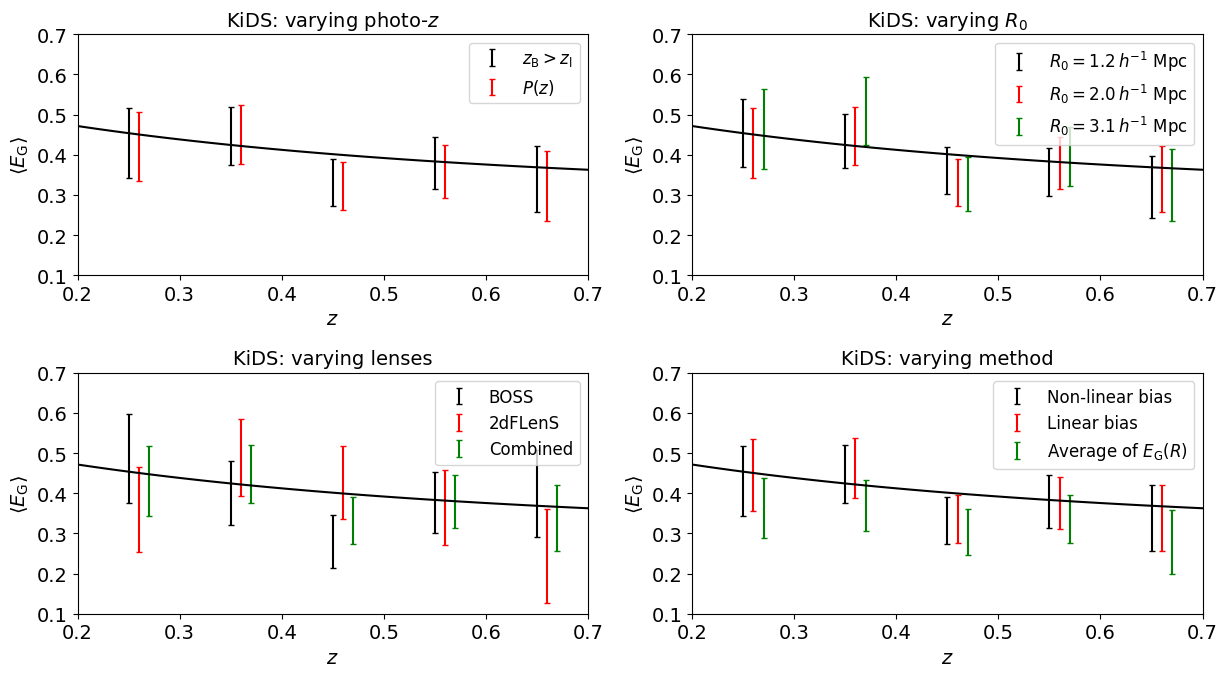}
\caption{Comparison of the scale-independent values $\langle
  E_{\mathrm{G}} \rangle$ determined for each LRG redshift slice
  correlated with the KiDS-1000 source sample, varying the
  galaxy-galaxy lensing analysis assumptions, sample and methodology.
  Our fiducial analysis adopted a point-based photo-$z$ correction
  with $z_{\mathrm{B}} > z_{\mathrm{l}}$, $R_0 = 2.0 \, h^{-1}$ Mpc
  and a model fitted to $(\Upsilon_{\mathrm{gm}},
  \Upsilon_{\mathrm{gg}})$ including non-linear galaxy bias.  The
  upper-left panel compares determinations of $\langle E_{\mathrm{G}}
  \rangle$ varying the photo-$z$ dilution correction method, the
  upper-right panel varies the small-scale parameter $R_0$, the
  lower-left panel varies the lens sample to consider BOSS and 2dFLenS
  separately, and the lower-right panel alters the fitting method to
  only include linear galaxy bias, and to use a direct fit to the
  $E_{\mathrm{G}}(R)$ values.  The model line in each case is the
  prediction $E_{\mathrm{G}}(z) = \Omega_{\mathrm{m}}/f(z)$, where
  $\Omega_{\mathrm{m}} = 0.3$.}
\label{figkidsegvsz}
\end{figure*}

We now consider the effect on our cosmological fits of varying our
fiducial analysis choices.  Fig.\ \ref{figkidsegvsz} is a compilation
of different determinations of $\langle E_{\mathrm{G}} \rangle$ in
each lens redshift slice.  The upper-left panel compares the fits
varying the photo-$z$ dilution correction for $\Delta\Sigma$ between
the point-based and distribution-based approaches, and the upper-right
panel shows measurements varying the small-scale parameter $R_0$.  The
lower-left panel considers separate determinations based on the BOSS
and 2dFLenS samples (using the BOSS clustering measurements in both
cases).  The lower-right panel alters the fitting method to only use a
linear-bias model (set $b_{\mathrm{NL}} = 0$), and to use a direct fit
to the $E_{\mathrm{G}}(R)$ measurements presented in the fifth row of
Fig.\ \ref{figstatkids}, as opposed to our fiducial fits to
$\Upsilon_{\mathrm{gm}}$ and $\Upsilon_{\mathrm{gg}}$.  In all cases,
the systematic variation of the recovered $E_{\mathrm{G}}$ values is
negligible compared to the statistical errors (noting that the BOSS
and 2dFLenS comparison is also subject to sample variance error).  We
find that varying the photo-$z$ dilution correction, choice of $R_0$
and fitting method produce a systematic variation of $0.08$-$\sigma$,
$0.24$-$\sigma$ and $0.38$-$\sigma$ respectively, when expressed as a
fraction of the statistical error $\sigma$.

\section{Summary}
\label{secsummary}

We have used the latest weak gravitational lensing data release from
the Kilo-Degree Survey, KiDS-1000, together with overlapping galaxy
redshift survey data from BOSS and 2dFLenS, to perform cosmological
tests associated with the relative amplitude of galaxy-galaxy lensing
and galaxy clustering statistics.  We quantified our results using the
$E_{\mathrm{G}}(R)$ statistic, which we were able to measure up to
projected separations of $100 \, h^{-1}$ Mpc, recovering a
scale-independent value of $\langle E_{\mathrm{G}} \rangle$ with
accuracies in the range 15-20\% in five lens redshift slices of width
$\Delta z = 0.1$.  The scale-dependence and redshift-dependence of
these measurements are consistent with the theoretical expectation of
general relativity in a Universe with matter density
$\Omega_{\mathrm{m}} \sim 0.3$.  The measurements are consistent with
a scale-independent model for $E_{\mathrm{G}}$, and constrain allowed
variation within $25\%$ (1-$\sigma$) error per decade in projected
scale.  Fitting our $E_{\mathrm{G}}$ dataset with a flat $\Lambda$CDM
model, we find $\Omega_{\mathrm{m}} = 0.27 \pm 0.04$.

We demonstrated that our results are robust against different analysis
methodologies.  In particular, we showed that:

\begin{itemize}

\item Source photometric redshift errors cause a significant dilution
  of the inferred projected mass density $\Delta\Sigma$, by causing
  unlensed foreground sources to appear in the background of lenses.
  We demonstrated that this dilution may be corrected either by
  modifying the estimator for $\Delta\Sigma$ to include the redshift
  probability distribution of the sources, or by using a source
  spectroscopic calibration sample to compute the multiplicative bias,
  producing consistent results.  When applied to the mock catalogue,
  these estimation methods recovered the $\Delta\Sigma$ measurement
  obtained using source spectroscopic redshifts.

\item A Gaussian analytical covariance for the galaxy-galaxy lensing
  statistics, with suitable modifications for the survey selection
  function and small-scale noise terms, predicted errors which agreed
  within $20\%$ with estimates from a jack-knife procedure and from
  the variation across mock realisations.

\item Our amplitude-ratio test, based on the annular differential
  density statistics $\Upsilon_{\mathrm{gm}}$ and
  $\Upsilon_{\mathrm{gg}}$, is insensitive to the small-scale
  parameter $R_0$ adopted in these statistics, producing consistent
  results for choices in the range $1 < R_0 < 3 \, h^{-1}$ Mpc.  We
  also obtained consistent results analysing BOSS and 2dFLenS
  separately.

\end{itemize}

We performed an additional series of tests jointly fitting an overall
amplitude $\sigma_8$ to the galaxy-galaxy lensing and clustering
statistics in our mocks, marginalising over linear and non-linear bias
parameters.  For our scenario, the projected galaxy-galaxy lensing
measurements produced a slightly more accurate determination of
$\sigma_8$ than the angular statistics after matching of scales,
although the results are fully consistent.

Our analysis sets the stage for upcoming, increasingly accurate,
cosmological tests using amplitude-ratio statistics, as gravitational
lensing and galaxy redshift samples continue to grow.  These datasets
will continue to offer rich possibilities for placing tighter
constraints on allowed gravitational physics.

\begin{acknowledgements}

We thank the anonymous referee for useful comments on the paper.  CB
is grateful to Rossana Ruggeri, Alexie Leauthaud, Johannes Lange and
Sukhdeep Singh for valuable discussions on galaxy-galaxy lensing
measurements. \\

We acknowledge support from: the Australian Research Council Centre of
Excellence for All-sky Astrophysics (CAASTRO) through project number
CE110001020 (CB, KG); the European Research Council under grant
numbers 647112 (MA, BG, CH, TT), 770935 (AD, HH, JLvdB, AHW) and
693024 (SJ); the Polish Ministry of Science and Higher Education
through grant DIR/WK/2018/12 and the Polish National Science Center
through grant 2018/30/E/ST9/00698 (MB); the Max Planck Society and the
Alexander von Humboldt Foundation in the framework of the Max
Planck-Humboldt Research Award endowed by the Federal Ministry of
Education and Research (CH); Heisenberg grant Hi 1495/5-1 of the
Deutsche Forschungsgemeinschaft (HH); the Beecroft Trust (SJ); Vici
grant 639.043.512 financed by the Netherlands Organisation for
Scientific Research (AK); the Alexander von Humboldt Foundation (KK);
the NSFC of China grant 11973070, the Shanghai Committee of Science
and Technology grant 19ZR1466600 and Key Research Program of Frontier
Sciences grant ZDBS-LY-7013 (HYS); and the European Union’s Horizon
2020 research and innovation programme under the Marie
Sk{l}odowska-Curie grant 797794 (TT). \\

The 2dFLenS survey is based on data acquired through the Australian
Astronomical Observatory, under program A/2014B/008. It would not have
been possible without the dedicated work of the staff of the AAO in
the development and support of the 2dF-AAOmega system, and the running
of the AAT. \\

Funding for SDSS-III has been provided by the Alfred P. Sloan
Foundation, the Participating Institutions, the National Science
Foundation, and the U.S. Department of Energy Office of Science. The
SDSS-III web site is http://www.sdss3.org/.  SDSS-III is managed by
the Astrophysical Research Consortium for the Participating
Institutions of the SDSS-III Collaboration including the University of
Arizona, the Brazilian Participation Group, Brookhaven National
Laboratory, Carnegie Mellon University, University of Florida, the
French Participation Group, the German Participation Group, Harvard
University, the Instituto de Astrofisica de Canarias, the Michigan
State/Notre Dame/JINA Participation Group, Johns Hopkins University,
Lawrence Berkeley National Laboratory, Max Planck Institute for
Astrophysics, Max Planck Institute for Extraterrestrial Physics, New
Mexico State University, New York University, Ohio State University,
Pennsylvania State University, University of Portsmouth, Princeton
University, the Spanish Participation Group, University of Tokyo,
University of Utah, Vanderbilt University, University of Virginia,
University of Washington, and Yale University. \\

We have used {\tt matplotlib} \citep{Hunter07} for the generation of
scientific plots, and this research also made use of {\tt astropy}, a
community-developed core Python package for Astronomy
\citep{Astropy13}. \\ \\

{\it Author contributions:} All authors contributed to the development
and writing of this paper. The authorship list is given in two groups:
the lead author (CB), followed by an alphabetical group who have
made a significant contribution to either the data products or to the
scientific analysis.

\end{acknowledgements}

\bibliographystyle{aa}
\bibliography{kids1000_ggl}

\begin{appendix}

\section{Covariance of average tangential shear}
\label{seccovgtap}

We may evaluate the covariance of $\gamma_{\mathrm{t}}$ between scales
$\theta$ and $\theta'$ using Eq.\ \ref{eqgtmod},
\begin{equation}
\begin{split}
  &
  \mathrm{Cov}[\gamma_{\mathrm{t}}^{ij}(\theta),\gamma_{\mathrm{t}}^{kl}(\theta')]
  = \\ & \int \frac{d^2\vl}{(2\mathrm{\pi})^2} \int
  \frac{d^2\vl'}{(2\mathrm{\pi})^2} \,
  \mathrm{Cov}[C_{\mathrm{g\kappa}}^{ij}(\vl) ,
    C_{\mathrm{g\kappa}}^{kl}(\vl')] \, J_2(\ell \theta) \, J_2(\ell'
  \theta') ,
\end{split}
\label{eqgtcovap}
\end{equation}
where $\gamma_{\mathrm{t}}^{ij}$ denotes the average tangential shear
of source sample $j$ around lens sample $i$.  We adopt an
approximation that different multipoles $\vl$ are uncorrelated such
that,
\begin{equation}
\mathrm{Cov}[C_{\mathrm{g\kappa}}^{ij}(\vl) ,
  C_{\mathrm{g\kappa}}^{kl}(\vl')] = \delta_{\mathrm{D}}(\vl - \vl')
\, \sigma^2(\ell) ,
\end{equation}
where $\delta_{\mathrm{D}}$ is the Dirac delta function, and the
variance $\sigma^2(\ell)$ is given by Eq.\ \ref{eqgtvar}.

Eq.\ \ref{eqgtcovap} becomes, after integrating the delta function $\int
d^2\vl' \, f(\vl') \, \delta_{\mathrm{D}}(\vl - \vl') =
\frac{(2\mathrm{\pi})^2}{\Omega} \, f(\vl)$ where $\Omega$ is the
total survey angular area in steradians,
\begin{equation}
  \mathrm{Cov}[\gamma_{\mathrm{t}}^{ij}(\theta),\gamma_{\mathrm{t}}^{kl}(\theta')]
  = \frac{1}{\Omega} \int \frac{d\ell \, \ell}{2\mathrm{\pi}} \,
  \sigma^2(\ell) \, J_2(\ell \theta) \, J_2(\ell \theta') .
\end{equation}
If the measurements of $\gamma_{\mathrm{t}}$ are averaged within
angular bins $m$ and $n$, where the angular area of the $i$th bin is
$\Omega_i$ (i.e.\ the area of the annulus between the bin limits),
then the covariance between the bins is,
\begin{equation}
\begin{split}
  C_{mn} &= \int_m \frac{d^2\theta}{\Omega_{\mathrm{m}}} \int_n
  \frac{d^2\theta'}{\Omega_n} \,
  \mathrm{Cov}[\gamma^{ij}_{\mathrm{t}}(\theta),\gamma^{kl}_{\mathrm{t}}(\theta')]
  \\ &= \frac{1}{\Omega} \int \frac{d\ell \, \ell}{2\mathrm{\pi}} \,
  \sigma^2(\ell) \, \overline{J_{2,m}}(\ell) \,
  \overline{J_{2,n}}(\ell) ,
\end{split}
\end{equation}
where $\overline{J_{2,n}}(\ell) = \int_{\theta_{1,n}}^{\theta_{2,n}}
\frac{2\mathrm{\pi}\theta \, d\theta}{\Omega_n} \, J_2(\ell \theta)$.
We also note that the contribution of any constant term in the
covariance $\sigma^2(\ell) = C$ (such as the noise terms) is,
\begin{equation}
\begin{split}
  C_{mn} &= C \, \int \frac{2\mathrm{\pi}\theta \,
    d\theta}{\Omega_{\mathrm{m}}} \int \frac{2\mathrm{\pi}\theta' \,
    d\theta'}{\Omega_n} \frac{1}{\Omega} \int \frac{d\ell \,
    \ell}{2\mathrm{\pi}} \, J_2(\ell \theta) \, J_2(\ell \theta')
  \\ &= \frac{C}{\Omega \, \Omega_n} \delta^{\mathrm{K}}_{mn} ,
\end{split}
\end{equation}
using the Bessel function relation $\int_0^\infty J_n(ax) \, J_n(bx)
\, x \, dx = \delta_{\mathrm{D}}(a-b)/b$.

\section{Modification of covariance for survey window}
\label{seccovwinap}

We derive how the covariance of a cross-correlation function between
two Gaussian fields, $\delta_1(\vx)$ and $\delta_2(\vx)$, is modified
by the window function of the fields, $W_1(\vx)$ and $W_2(\vx)$.  We
adopt the case of a 2D flat sky, where the vector separation $\vr$
between two points has magnitude $r$ and orientation angle $\theta$.
An estimator of the cross-correlation function of the fields at
separation $r$ is,
\begin{equation}
\hxi(r) = \frac{1}{A_2(r)} \int \frac{d\theta}{2\mathrm{\pi}} \int
d^2\vx \, \delta_1(\vx) \, \delta_2(\vx+\vr) \, W_1(\vx) \,
W_2(\vx+\vr) ,
\end{equation}
where $A_2(r) = \int \frac{d\theta}{2\mathrm{\pi}} \int d^2\vx \,
W_1(\vx) \, W_2(\vx+\vr)$.  The expectation value of this expression
is,
\begin{equation}
\begin{split}
  \langle \hxi \rangle &= \frac{1}{A_2(r)} \int
  \frac{d\theta}{2\mathrm{\pi}} \int d^2\vx \langle \delta_1(\vx)
  \delta_2(\vx+\vr) \rangle W_1(\vx) W_2(\vx+\vr) \\ &=
  \frac{1}{A_2(r)} \int \frac{d^2\vk}{(2\mathrm{\pi})^2} \,
  P_{12}(\vk) \int \frac{d\theta}{2\mathrm{\pi}} \int d^2\vx W_1(\vx)
  W_2(\vx+\vr) \mathrm{e}^{-\mathrm{i}\vk\cdot\vr} \\ &\approx
  \frac{1}{A_2(r)} \frac{1}{2\mathrm{\pi}} \int dk \, k \, P_{12}(k)
  \, \int \frac{d\theta}{2\mathrm{\pi}} \, A_2(r) \,
  \mathrm{e}^{-\mathrm{i}kr\cos{\theta}} \\ &= \frac{1}{2\mathrm{\pi}}
  \int dk \, k \, P_{12}(k) \, J_0(kr) ,
\end{split}
\label{eqxiwin}
\end{equation}
where we have introduced the cross-power spectrum $P_{12}(k)$, and the
approximation in the third line of Eq.\ \ref{eqxiwin} ignores the
$\theta$ dependence of $\int d^2\vx \, W_1(\vx) \, W_2(\vx+\vr)$.

The covariance of the estimator may be deduced from,
\begin{equation}
\begin{split}
  & \langle \hxi(\vr) \, \hxi(\vs) \rangle = \frac{1}{A_2(r) \,
    A_2(s)} \int d^2\vx \int d^2\vy \, \\ & A_{12}(\vx,\vr) \,
  A_{12}(\vy,\vs) \, \langle \delta_1(\vx) \, \delta_2(\vx+\vr) \,
  \delta_1(\vy) \, \delta_2(\vy+\vs) \rangle ,
\end{split}
\end{equation}
where we have written $A_{12}(\vx,\vr) = W_1(\vx) \, W_2(\vx+\vr)$.
Expanding this expression using Wick's theorem for a Gaussian random
field, $\langle \delta_1 \, \delta_2 \, \delta_3 \, \delta_4 \rangle =
\langle \delta_1 \, \delta_2 \rangle \langle \delta_3 \, \delta_4
\rangle + \langle \delta_1 \, \delta_3 \rangle \langle \delta_2 \,
\delta_4 \rangle + \langle \delta_1 \, \delta_4 \rangle \langle
\delta_2 \, \delta_3 \rangle$, we find
\begin{equation}
\begin{split}
& \mathrm{Cov}[\hxi(\vr),\hxi(\vs)] = \frac{1}{A_2(r) \, A_2(s)} \int
  d^2\vx \int d^2\vy \, \\ & A_{12}(\vx,\vr) \, A_{12}(\vy,\vs) \, [
    \langle \delta_1(\vx) \, \delta_1(\vy) \rangle \langle
    \delta_2(\vx+\vr) \, \delta_2(\vy+\vs) \rangle \\ & + \langle
    \delta_1(\vx) \, \delta_2(\vy+\vs) \rangle \langle
    \delta_2(\vx+\vr) \, \delta_1(\vy) \rangle ] .
\end{split}
\end{equation}
Using $\langle \delta_i(\vx) \, \delta_j(\vy) \rangle = \langle
\delta_i(\vx) \, \delta^*_j(\vy) \rangle = \int
\frac{d^2\vk}{(2\mathrm{\pi})^2} \, P_{ij}(\vk) \,
\mathrm{e}^{-\mathrm{i}\vk\cdot(\vx-\vy)}$, and omitting some algebra,
the first term evaluates to,
\begin{equation}
  \int \frac{d^2\vk}{(2\mathrm{\pi})^2} \, P_{11}(\vk) \, P_{22}(\vk)
  \, \mathrm{e}^{\mathrm{i}\vk\cdot(\vr-\vs)} \int d^2\vx \,
  A_{12}(\vx,\vr) \, A_{12}(\vx,\vs) ,
\end{equation}
and the second term evaluates to,
\begin{equation}
  \int \frac{d^2\vk}{(2\mathrm{\pi})^2} \, P^2_{12}(\vk) \,
  \mathrm{e}^{\mathrm{i}\vk\cdot(\vr-\vs)} \int d^2\vx \,
  A_{12}(\vx,\vr) \, A_{12}(\vx,\vs) .
\end{equation}
The expression for the covariance is then,
\begin{equation}
\begin{split}
 &\mathrm{Cov}[\hxi(\vr),\hxi(\vs)] = \frac{\int d^2\vx \,
    A_{12}(\vx,\vr) \, A_{12}(\vx,\vs)}{A_2(r) \, A_2(s)} \\ &\int
  \frac{d^2\vk}{(2\mathrm{\pi})^2} \, \left[ P_{11}(\vk) \,
    P_{22}(\vk) + P^2_{12}(\vk) \right] \,
  \mathrm{e}^{\mathrm{i}\vk\cdot(\vr-\vs)} .
\end{split}
\end{equation}
Averaging the estimator over angles we obtain,
\begin{equation}
\begin{split}
 & \mathrm{Cov}[\hxi(r),\hxi(s)] \approx \frac{A_3(r,s)}{A_2(r) \,
    A_2(s)} \\ & \frac{1}{2\mathrm{\pi}} \int dk \, k \, \left[
    P_{11}(k) \, P_{22}(k) + P^2_{12}(k) \right] \, J_0(kr) \, J_0(ks)
  ,
\end{split}
\end{equation}
where $A_3(r,s) = \int d^3\vr \int d^3\vs \int d^2\vx \,
A_{12}(\vx,\vr) \, A_{12}(\vx,\vs)$.

\section{Combining correlated tomographic slices}
\label{seccombap}

In order to reduce the size of a data vector, we can optimally combine
separate correlated estimates of a statistic, such as a galaxy-galaxy
lensing measurement for a given lens sample against different
tomographic source slices.  This procedure is an example of data
compression \citep{Tegmark97}.

Suppose we have measured a given statistic at $N_{\mathrm{r}}$
different scales, for $N_{\mathrm{s}}$ source tomographic slices, and
we wish to average the statistic over source samples, where the
measurements in the different slices are correlated.  We'll arrange
these quantities in a data vector $\mathbf{x}$ of length
$N_{\mathrm{s}} N_{\mathrm{r}}$ with corresponding covariance matrix
$\mathbf{C}$ of dimension $N_{\mathrm{s}} N_{\mathrm{r}} \times
N_{\mathrm{s}} N_{\mathrm{r}}$.  The operation to combine the
different tomographic slices to a compressed data vector $\mathbf{y}$
of length $N_{\mathrm{r}}$ can be written as,
\begin{equation}
  \mathbf{y} = \mathbf{w}^{\mathrm{T}} \mathbf{x} ,
\end{equation}
where $\mathbf{w}$ is a weight matrix of dimension $N_{\mathrm{s}}
N_{\mathrm{r}} \times N_{\mathrm{r}}$, and we normalise the weights
such that the column corresponding to each scale bin sums to unity.
The optimal choice of weight matrix \citep{Tegmark97} is,
\begin{equation}
  \mathbf{w} = \mathbf{C}^{-1} \, \mathbf{D} ,
\end{equation}
where $\mathbf{D}$ is a matrix of dimension $N_{\mathrm{s}}
N_{\mathrm{r}} \times N_{\mathrm{r}}$, whose columns consist of
$N_{\mathrm{s}} N_{\mathrm{r}}$ entries for each final scale bin, with
value 1 when the entry in $\mathbf{x}$ corresponds to the same scale
bin, and value 0 otherwise.  The resulting covariance matrix of
$\mathbf{y}$ is,
\begin{equation}
  \mathbf{C}_{\mathrm{y}} = \mathbf{w}^{\mathrm{T}} \, \mathbf{C} \,
  \mathbf{w} ,
\label{eqcovcompress}
\end{equation}
which has dimension $N_{\mathrm{r}} \times N_{\mathrm{r}}$.

We found that this data compression scheme is more robust against
numerical issues with the matrix inverse (and suffers negligible loss
in precision) if we replaced the weight matrix with $\mathbf{w} =
\mathbf{V}^{-1} \, \mathbf{D}$, where $\mathbf{V}$ is a diagonal
matrix just containing the variance of the measurements.  In this
implementation the weight matrix is slightly suboptimal, but we
retained the full covariance matrix in Eq.\ \ref{eqcovcompress} to
ensure correct error propagation.

\end{appendix}

\end{document}